\documentclass[fleqn,usenatbib]{mnras}

\usepackage{graphicx}	
\usepackage{amsmath}	
\usepackage{amssymb}	
\usepackage{multicol}        
\usepackage{bm}		
\usepackage{pdflscape}	
\usepackage{dcolumn}
\usepackage{xcolor}
\usepackage{mathtools}
\usepackage[T1]{fontenc}
\usepackage{ae,aecompl}
\usepackage{hyperref}
\usepackage{natbib}
\usepackage[normalem]{ulem}
\usepackage{fontawesome}
\usepackage{newtxtext}

\newcommand{\redmagic}{\texttt{redMaGiC} }
\newcommand{\bq}{\boldsymbol q}
\newcommand{\bx}{\boldsymbol x}
\newcommand{\bk}{\textbf{k}}
\newcommand{\bPsi}{\boldsymbol{\Psi}}

\newcommand{\ihmpc}{\,h{\rm Mpc}^{-1}}
\newcommand{\aemulus}{\texttt{Aemulus} }
\newcommand{\kmax}{k_{\rm max}}
\author[Kokron et al.]{Nickolas Kokron$^{1,2}$
\thanks{Contact e-mail: \href{mailto:kokron@stanford.edu}{kokron@stanford.edu}}%
, Joseph DeRose$^{3}$, Shi-Fan Chen$^{3,4}$, Martin White$^{3,4,5}$, Risa H. Wechsler$^{1,2}$
\\
$^{1}$ Kavli Institute for Particle Astrophysics and Cosmology and Department of Physics, Stanford University, 382 Via Pueblo Mall, Stanford, CA 94305, USA \\ 
$^{2}$ Kavli Institute for Particle Astrophysics and Cosmology, SLAC National Accelerator Laboratory, 2575 Sand Hill Road, Menlo Park, CA 94025, USA \\
$^{3}$ Physics Division, Lawrence Berkeley National Laboratory, Berkeley, CA 94720\\
$^{4}$ Department of Physics, University of California, Berkeley, CA 94720 \\
$^{5}$ Department of Astronomy, University of California, Berkeley, CA 94720
}
\pubyear{2021}
\date{}
\title[HOD Stochasticity]{Priors on red galaxy stochasticity from hybrid effective field theory}

\begin{document}

\maketitle


\begin{abstract}
We investigate the stochastic properties of typical red galaxy samples in a controlled numerical environment. We use Halo Occupation Distribution (HOD) modelling to create mock realizations of three separate bright red galaxy samples consistent with datasets used for clustering and lensing analyses in modern galaxy surveys. Second-order Hybrid Effective Field Theory (HEFT) is used as a field-level forward model to describe the full statistical distribution of these tracer samples, and their stochastic power spectra are directly measured and compared to the Poisson shot-noise prediction. While all of the galaxy samples we consider are hosted within haloes with sub-Poisson stochasticity, we observe that the galaxy samples themselves possess stochasticities that range from sub-Poisson to super-Poisson, in agreement with predictions from the halo model. As an application of our methodology, we place priors on the expected degree of non-Poisson stochasticity in cosmological analyses using such samples. We expect these priors will be useful in reducing the complexity of the full parameter space for future analyses using second-order Lagrangian bias models. More generally, the techniques outlined here present the first application of hybrid EFT methods to characterize models of the galaxy--halo connection at the field level, revealing new connections between once-disparate modelling frameworks.

\end{abstract}
\maketitle

\section{Introduction}
\label{sec:intro}
Spectroscopic galaxy redshift surveys are poised to produce some of the leading cosmological datasets of the upcoming decade. The Dark Energy Spectroscopic Instrument (DESI, \citealt{Aghamousa:2016zmz}), for example, will observe an order of magnitude more galaxies than the incredibly successful Sloan Digital Sky Surveys \citep{Dawson_2012,Dawson_2016}. Other galaxy survey probes, such as the Vera Rubin Observatory's Legacy Survey of Space and Time (LSST, \citet{Ivezic:2008fe,Mandelbaum:2018ouv}), will measure the shapes of roughly ten billion of galaxies and tease out the correlated weak gravitational lensing signal therein, directly measuring the gravitational effect of dark matter at unprecedented statistical power over half the sky. The upcoming increase of statistical power afforded by next-generation cosmic surveys is especially timely, as they will shed light on several `tensions' that have crept up between cosmological datasets over the recent years. These include the tension over measurements of the Hubble constant \citep[or, alternatively, in the sound horizon]{Di_Valentino_2021} between early and late-Universe probes and the recently hinted $S_8$ tension over the amplitude of density fluctuations in the Universe compared to those predicted by observations of the cosmic microwave background (CMB) and the assumption of $\Lambda$CDM \citep{krolewski2021cosmological, descollaboration2021dark, Heymans_2021,white2021cosmological}. Such tensions could signify a breakdown of the standard cosmological model, $\Lambda$CDM, if validated by larger datasets. 
\par 
However, the vast statistical power of future datasets simultaneously presents a significant challenge to the models we use to analyze them. Significant care must be taken to ensure that their accuracy is sub-dominant compared to statistical and systematic uncertainties; mischaracterizing the accuracy of models for describing the statistical properties of galaxy surveys could lead to biased inferences on the properties of the Universe. \par 
In particular, models for the statistical properties of galaxy distributions must surmount two individual challenges. First, the statistical properties of the late-time dark matter distribution itself, given a cosmological model, must be well understood. This is a challenging task, as the gravitational collapse problem of the cold dark matter fluid is a non-linear process. Significant progress in this regard has been made via the numerical study of this problem, using $N$-body simulations of structure formation \citep{hockney1988computer, Bagla:2004au,Kuhlen:2012ft,Schneider_2016}. Despite the large computational cost of running individual $N$-body simulations, computational power and statistical tools have progressed significantly, and one can now run suites of simulations that span several points in cosmological parameter space. With these suites, \emph{emulators} have become commonplace tools in predicting the non-linear properties of structure formation. Measurements of any given observable across the simulation suite serve as inputs to models that predict non-linear statistics of the dark matter distribution rapidly, and their accuracy can be well calibrated given a suitable experimental design \citep{Heitmann_2013,Garrison_2018,DeRose:2018xdj,Knabenhans:2018cng,angulo2021bacco}\par 
An emulator for dark matter statistics alone, however, still cannot be used to predict signals of the clustering properties of galaxies. Modelling the \emph{galaxy--halo connection}, or more broadly, the \emph{tracer--matter connection} is the second challenge that must be surmounted in order to construct a model suitable for end-to-end analysis of galaxy survey data. Models for the tracer--matter connection fall into several different categories. Here, we highlight two such categories: empirical/statistical and analytic/perturbative models of the tracer--matter connection.\par 
Empirical models include so-called halo occupation distributions (HOD), (sub)-halo abundance matching, and direct modeling of the formation histories of galaxies, among others. Empirical models attempt to infer statistical relations between halos identified in dark matter simulations and mock galaxy populations that inhabit them, in light of both observational data on the given population and properties of the host dark matter haloes. Such data includes, for example: two-and-three point correlation functions \citep{Zheng:2004id, Yuan:2018qek}, luminosity functions \citep{yang2003, cooray2006}, and measurements of stellar mass functions and star formation rates \citep{Behroozi_2013}. Empirical models allow for deep insights into galaxy formation and evolution \citep{behroozi2019}, the creation of realistic mock realizations of sky surveys \citep{wechsler2021addgals} as well as offer potent frameworks to describe the statistical properties of galaxies down to very small scales \citep{derose2021modeling}. We refer the readers to \cite{Wechsler:2018pic} for a comprehensive review on empirical models and other simulation-based models of the galaxy--halo connection.\par 
Perturbative models for the tracer--matter connection, also known as \emph{bias models} (see \citealt{Desjacques:2016bnm} for a comprehensive review), try to capture the relationship between the dark matter and a population of tracers in a different way. Instead of explicitly relating the properties of haloes to the properties of galaxy samples, bias models specify a functional form for the relation between the large-scale, smoothed, dark matter density and the density of tracers under consideration. This functional form is restricted by a set of symmetries that hold in the relation, and the given order in powers of the aforementioned density one is working in. Each term in the bias expansion is accompanied by a free coefficient that captures the response of the tracer population to that term. The flexible parameterization of bias models imply they should be able to describe, within their regime of applicability, the statistical properties of \emph{any} tracer sample whose properties obey the imposed symmetries.\par 
While the bias expansion captures the deterministic relation between a tracer's distribution and that of the underlying matter field, there is an additional stochastic component in this relation that decorrelates the two fields, due to small-scale processes. This noise, scatter, or \emph{stochasticity} contribution to the tracer--matter connection is an important component that must be understood if we wish to extract the most information from our datasets. Notably, stochasticity becomes important at small scales where higher-order bias terms are also expected to be significant, and their impact on observables is partially degenerate with these bias terms. Indeed, stochasticity has previously been defined as not only this random component but also as the impact of not including higher-order bias terms in a model for the distribution of galaxies \citep{Baldauf_2013}. A lack of prior understanding of the effects of stochasticity can lead to significant degradation of cosmological constraining power. Significant efforts have been undertaken to characterize the stochastic properties of galaxy samples, however not to the same extent as galaxy bias. We refer the reader to \citet{Baldauf_2016,Paech_2017,Ginzburg_2017,friedrich2021pdf,Sullivan21} for previous discussions on the interplay between stochasticity and bias modelling. \par 
In this work we use field-level realizations of Lagrangian bias models with fully non-linear dark matter dynamics, an approach recently dubbed Hybrid Effective Field Theory (HEFT), to study the properties of specfic galaxy samples, focusing on the example red galaxy samples that are used as both clustering and lens samples in cross-correlation analyses. Specifically, we use HEFT to measure the amplitude of the stochasticity of said samples, using fits to their properties via HODs as a proxy for their statistical properties. These measurements can then be used to place priors on subsequent analyses that help reduce the computational complexity of the inference procedure. Our methodology highlights the synergies in using multiple models of the tracer--matter connection to study the same galaxy sample. The study of stochasticity we describe below also raises a number of new ways that bias models and empirical models may be combined to shed light into the ways that galaxies, or haloes, populate the broader large-scale structure of the Universe. \par 
The paper is structured as follows: in \S~\ref{subsec:HOD}, \ref{subsec:HEFT} we give a brief overview of both HODs and hybrid EFT, the two statistical tools we use to characterize the tracer--matter connection in this paper. In \S~\ref{subsec:Perrstimate} we outline our procedure to use field-level realizations of the bias model to both estimate the bias parameters of HOD samples and consequently measure their stochastic power spectra. In \S~\ref{subsection:analyticnonpoiss} we review some results on the causes of non-Poisson stochasticity in the framework of the halo model. Specifically we discuss two competing effects, one-halo \emph{enhancement} and halo \emph{exclusion}, which drive the large-scale stochasticity of galaxy samples to the super- and sub-Poisson regime, respectively. In \S~\ref{sec:simsnsamples} we outline both the simulation suite used and the mock galaxies we populate onto these simulations using said HODs. We discuss the functional forms used and the derived parameters of the HODs for the three samples under consideration, as well as how their differences frame our results. In \S~\ref{sec:results} we report the results of our stochastic power spectrum measurement procedure applied to halo samples and to HOD samples. Specifically, in \S~\ref{subsec:haloperr} we look at previous results in the literature on the stochasticity of haloes within the context of our HEFT model. In \S~\ref{subsec:hodperr} we report the results of the same procedure on our HOD samples, as well as a discussion and some interpretation of the results that we find. In \S~\ref{subsec:chainhod} we address a related question of the distribution of large-scale stochasticities for all HOD models that are consistent with a given sample of DESI-like luminous red galaxies (LRGs). In \S~\ref{subsec:priors} we present priors on the allowed range of deviations from Poisson stochasticity that we expect, as a result of our experiments across different samples of galaxies and within the DESI-like sample.
\section{Methods}
\label{sec:methods}
\begin{figure*}
    \centering
    \includegraphics[width=\textwidth]{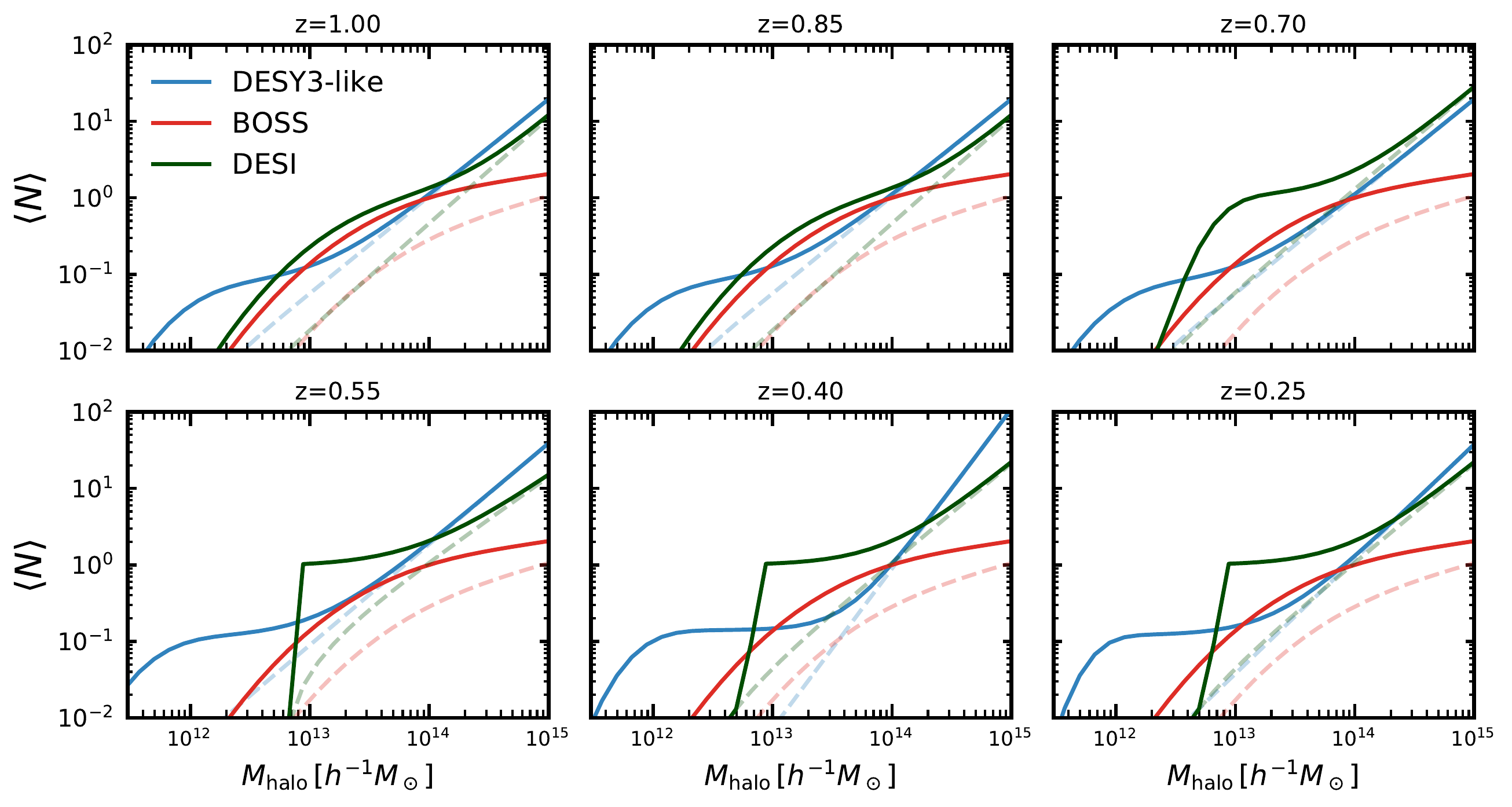}
    \caption{Mean HOD (average number of galaxies as a function of halo mass) for three red galaxy samples considered in this work. The color represents the type of HOD adopted, with solid (dashed) lines showing the number of total (satellite) galaxies per halo mass, respectively. }
    \label{fig:hodplots}
\end{figure*}

\subsection{Halo Occupation Distributions}
\label{subsec:HOD}
Halo occupation distribution modelling is an empirical parameterization of the way galaxies occupy haloes. This is done via a probabilistic mapping that specifies the average number of galaxies of a kind that are hosted within a halo. The standard HOD models \citep{Berlind_2002,Zheng:2004id,Zheng:2007zg} separate these galaxies into \emph{central} and \emph{satellite} galaxies. A commonly used parameterization is given by 
\begin{equation}
\label{eqn:ncen}
    \langle N_{\rm cen} (M) \rangle = \frac{f_{\rm cen}}{2} \left [ 1 + {\rm erf} \left (\frac{\log M - \log M_{\rm min}}{\sigma_{\log M}} \right ) \right ],
\end{equation}
and 
\begin{equation}
\label{eqn:nsat}
    \langle N_{\rm sat} (M) \rangle = \left [ \frac{M - M_0}{M_1} \right]^\alpha.
\end{equation}
This totals six parameters, however the parameter $f_{\rm cen}$ is usually fixed to $f_{\rm cen} = 1$. Additionally, some HOD models adopting this parameterization also alternate between using $\langle N_{\rm sat} (M)\rangle$ and $\langle N_{\rm sat} (M) \rangle\langle N_{\rm cen} (M) \rangle$ for the expected number of satellites \citep[see e.g. discussion in][]{reddick2013}. Physically, this corresponds to down-weighting systems without centrals as also being less likely to host satellite galaxies. \par 
While the parameterization of Eqns.~\ref{eqn:ncen} and ~\ref{eqn:nsat} are standard, it is by no means an exhaustive list of occupation prescriptions adopted in the literature. First due to only depending on halo mass they are inevitably incomplete, as it is known that properties other than halo mass  (such as concentration, spin, and local environment) influence summary statistics \citep{Wechsler:2001cs, Gao:2005ca, Wechsler:2005gb,Dalal:2008zd,Mao:2017aym,Salcedo_2018,Chue_2018,Mansfield:2019ter}. This phenomenon, known as assembly bias, has motivated extensions to the standard HOD \citep{Hearin_2016, Yuan:2018qek}. These extensions can capture the dependence of halo occupation on additional properties beyond mass, but at the cost of introducing more free parameters. \par 
Nevertheless, it seems that for samples of LRGs observed by current and upcoming surveys this parameterization might be sufficient \citep[however, see \citet{yuan2021abacushod} for further discussion on this topic]{zacharegkas2021dark}. Understanding the applicability of the standard HOD parameterization is an active subject of research \citep{Hadzhiyska_2020}, with recent results indicating that alternative samples of galaxies such as emission line galaxies observed by DESI will require alternative parameterizations \citep{Hadzhiyska_2021}.
\par 

\subsection{Hybrid Effective Field Theory}
\label{subsec:HEFT}
Lagrangian biasing theory establishes a statistical relationship between the smoothed, large-scale properties of any given tracer sample and the underlying matter distribution at very high redshift \citep{Matsubara2008}. The functional form of this expression is given by a functional series expansion in the quantities allowed by the equivalence principle, rotational symmetry and translational symmetry at the Lagrangian coordinates $\bq$:
\begin{equation}
\label{eqn:lagbias}
    \delta_h (\bq) = F[\delta (\bq), s_{ij}(\bq)] + \epsilon(\bq),
\end{equation}
where $\delta_h$ is the proto-tracer density contrast, $\delta$ is the matter density contrast, and $s_{ij}$ is the traceless tidal tensor field. The field $\epsilon(\bq)$ is a stochastic field that captures the fact the process of tracer formation is not purely deterministic when considering the large-scale smoothed fields that this expansion is applicable to. Including all terms to second order, the expansion of $F[\delta(\bq), s_{ij}(\bq)]$ is given by \citep{Vlah_2016}
\begin{align}
\label{eqn:secondorder}
F[\delta (\bq), s_{ij}(\bq)]  \approx\, &1 + b_1 \delta (\boldsymbol{q}) + b_2 (\delta^2(\boldsymbol{q}) - \langle \delta^2 \rangle )\,+\\
\nonumber & b_{s^2} (s^2(\boldsymbol{q}) - \langle s^2 \rangle)+ \,b_{\nabla^2}\nabla^2 \delta(\boldsymbol{q}) .
\end{align}
At low redshifts, the statistical properties of these tracer fields depend on the combination of the initial relation of Eqn.~\ref{eqn:lagbias} and the time evolution of its ingredients under the influence of gravity. This time evolution is captured by the advection process from the Lagrangian coordinates to the late-time positions of tracer particles of the matter density
\begin{equation}
    \bx = \bq + \bPsi (\bq),
\end{equation}
where $\bPsi(\bq)$ is the Lagrangian displacement vector. Under number density conservation, the distribution of tracers at late times is then given by 
\begin{align}
    1 + \delta_h (\bx) = \int d^3 q &\left [ F [\delta (\bq), s_{ij}(\bq)] + \epsilon (\bq)\right ] \\
    \nonumber &\times \delta^D \left ( \bx - \bq - \bPsi (\bq) \right ).
\end{align}
Using Lagrangian Perturbation Theory (LPT), one can analytically determine order-by-order the properties of $\bPsi(\bq)$ \citep{Matsubara_2008, Carlson_2012}. These can then be combined with the expansion in Eqn.~\ref{eqn:secondorder} to create a model for the summary statistics of $\delta_h(\bq)$ (see \citealt{Chen_2020, Chen_2021} and references therein for an overview of the state of LPT). \par 
It has recently been pointed out \citep{modichenwhite19} that $N$-body simulations of structure formation similarly solve for $\bPsi(\bq)$, however in a non-perturbative way. This then implies that the ingredients of the bias expansion can be combined with the numerical displacements from an $N$-body simulation to create late-time representations of the basis fields that compose the expansion. \par 
At late times, a tracer field under this second-order hybrid expansion is explicitly written as 
\begin{align}
    \label{eqn:latetime}
    \delta_{h} (\bx) =& \delta_m(\bx) + b_1 \mathcal{O}_\delta (\bx) + b_{\nabla^2} \mathcal{O}_{\nabla^2 \delta} (\bx) + \\
    &b_2 \mathcal{O}_{\delta^2} (\bx) + b_{s^2} \mathcal{O}_{s^2} (\bx) + \epsilon (\bx),
\end{align}
where $\delta_m(\bx)$ is the dark matter density contrast from the simulation. The operators $\mathcal{O}_i$ are constructed by advecting each operator to late-times using the $\bPsi(\bq)$ obtained from simulations. 
\par
This combination of analytic bias and $N$-body displacements forms the model known as Hybrid Effective Field Theory (HEFT) and it is the main tool used in this paper to explore the properties of the $\epsilon (\bx)$ stochastic field. We refer the reader to \citet{modichenwhite19, Kokron_2021, zennaro2021bacco} for further discussions of HEFT and to \citet{hadzhiyska2021hefty} for an application of HEFT to survey data. \par 
\subsection{Bias parameters and stochastic spectra}
\label{subsec:Perrstimate}
From our construction of the field-level model for the tracer overdensity (Eqn.~\ref{eqn:latetime}), we can rearrange the expression to provide an estimate of the stochastic field $\epsilon (\bk)$ for a given set of bias parameters $b_i$
\begin{equation}
\label{eqn:epsilon}
    \epsilon (\bk) = \delta_h (\bk) - \delta_m (\bk) - \sum_i b_i \mathcal{O}_i (\bk),
\end{equation}
where the variable $\bk$ indicates that we are treating the fields in Fourier space. The \emph{stochastic power spectrum} is then defined as the power spectrum of this residual field for a given realization\footnote{A similar quantity was the object of study in \cite{Hamaus_2010} and \cite{Baldauf_2013}, however there the `stochastic field' was defined in the Eulerian frame explicitly as $\epsilon (\bk) = \delta_h (\bk) - b_1 \delta_m (\bk)$.} 
\begin{equation}
    P_{\rm err} (k) \equiv \langle \epsilon(\bk) \epsilon (-\bk) \rangle.
\end{equation}

We can use Eqn.~\ref{eqn:epsilon} and the standard estimator for the expectation value in the power spectrum to find an explicit expression for $P_{\rm err}(k)$ as a function of the bias parameters 
\begin{equation}
    P_{\rm err} (k) = \frac{1}{N_{\rm }(k)} \sum_{\bk \in \mathcal{S}(k)} \left \| \delta_h (\bk) - \delta_m (\bk) - \sum_i b_i \mathcal{O}_i (\bk) \right \|^2,
\end{equation}
where $\mathcal{S}(k)$ is the spherical shell of radius $k$ and width $dk$ and $N(k)$ is the number of Fourier modes that fall within $\mathcal{S}(k)$. \par 
Previous studies have estimated the best-fit bias parameters at a given scale by solving the least-squares problem of minimizing the error power spectrum \citep{Schmittfull_2019}, leading to the so-called \emph{bias transfer functions}:
\begin{equation}
\label{eqn:bhatold}
    \hat{b}_i (k) = \langle \mathcal{O}_i \mathcal{O}_j \rangle^{-1}(k) \left \langle \mathcal{O}_j(-\bk) \left [ \delta_h(\bk)  - \delta_m(\bk)  \right]\right \rangle.
\end{equation}
If our model for the tracer field is sufficiently accurate, then $P_{\rm err}(k)$ should correspond solely to the power spectrum of stochastic contributions. However, the determinations at a given $k$-scale are independent and one could find that the estimate of $\hat{b}_i(k_*)$ with $k_*$ a small-scale mode could degrade the fit to the error power spectrum at large-scales. Instead, we seek a comparable estimator for these bias parameters that appropriately penalizes over-fitting at small scales. \par 
We first apply a low-pass sharp-$k$ filter to $\epsilon(\bx)$ to remove the influence of very small scale modes in finding the optimal bias parameters. We represent these smoothed fields as $[\epsilon(\bx)]_{k_{\rm max}}$. If we then choose to minimize the average configuration-space stochastic field squared 
\begin{equation}
    \label{eqn:lossfunc}
    S = \langle [\epsilon(\bx)]_{k_{\rm max}}^2 \rangle,
\end{equation}
we find a loss function that is very similar to the \emph{EFT likelihood} of \cite{Schmidt_2019} and \cite{Cabass_2020}:
\begin{equation}
    S  \approx \int_{|\bk| < k_{\rm max}} \frac{d^3k}{(2\pi)^3}  \left \| \delta_h (\bk) - \delta_m (\bk) - \sum_i b_i \mathcal{O}_i (\bk) \right \|^2 .
\end{equation}
Minimizing $S$ with respect to bias parameters leads to an estimator $\hat{b}_i$ comparable to Eqn.~\ref{eqn:bhatold} but that includes information from \emph{all} modes until a maximum $k_{\rm max}$
\begin{equation}
\label{eqn:bhat}
    \hat{b}_i = M^{-1}_{ij} A_j, \\
\end{equation}
where $A_j$ and $M_{ij}$ are defined as 
\begin{align}
\label{eqn:aj}
    A_j &= \langle [\mathcal{O}_j(\bx)\left (\delta_h (\bx) - \delta_m (\bx) \right)]_{k_{\rm max}}\rangle, \\
     &= \int\limits_{\mathclap{|\bk| < k_{\rm max}}} \frac{d^3k}{(2\pi)^3} \mathcal{O}_j (\bk) [\delta_h - \delta_m]^*(\bk), 
\end{align}
and
\begin{align}
\label{eqn:mij}
    M_{ij} &= \langle [\mathcal{O}_i (\bx) \mathcal{O}_j (\bx)]_{k_{\rm max}} \rangle,\\
    &=\int \limits_{\mathclap{|\bk| < k_{\rm max}}}\frac{d^3k}{(2\pi)^3} \mathcal{O}_i (\bk) \mathcal{O}_j^*(\bk).
\end{align}

The procedure we adopt to estimate $P_{\rm err}(k)$ is then as follows.  We obtain estimates for the bias parameters using Eqn.~\ref{eqn:bhat}. We proceed to use these $\hat{b}_i $ to create realizations of the tracer fields and subtract these from the tracer sample, realizing Eqn.~\ref{eqn:epsilon} and our estimate $\hat{\epsilon}(\bk)$
\begin{equation}
\label{eqn:epsilonhat}
    \hat{\epsilon} (\bk) = \delta_h (\bk) - \delta_m (\bk) - \sum_i \hat{b}_i  \mathcal{O}_i (\bk).
\end{equation}
The stochastic power spectrum is then estimated directly from the fields constructed with Eqn.~\ref{eqn:epsilonhat}. Our fiducial figures are made adopting $k_{\rm max} = 0.4 \ihmpc$, but in Appendix ~\ref{appendix:B} we show the impact of varying $k_{\rm max}$ on a subset of our results.  \par 
The standard parameterization for this solely stochastic contribution can be informed by the symmetries of the occupation procedure and has a form broadly given by the power series \citep{Desjacques:2016bnm, Cabass_2020}
\begin{equation}
\label{eqn:poisson}
    P_{\rm err} (k) = \frac{1}{\bar{n}} \left [ a_1 + a_2 k^2 + \cdots \right],
\end{equation}
where $\bar{n}$ is the number density of the tracer sample in question. If the stochasticity of the sample arises solely due to Poisson statistics, then this corresponds to $a_1 = 1$ and $a_{2,3,\cdots} = 0$. This is the standard \emph{Poisson shot-noise} form of the stochastic power spectrum. As noted in \cite{Baldauf_2013} the observed form can differ from the standard Poisson form of Eq.~\ref{eqn:poisson} due to halo exclusion and non-linearities in clustering. We will return to this point shortly.\par 
The estimate of $\hat{b}_i$, along with our assumption that the fundamental field-level parameters are constant, may also be used to assess the validity of the model. At the scale $k_M$ where the bias expansion is assumed to break down we would expect the estimates to begin running strongly with scale, as well as using $k_{\rm max} > k_M$ in estimating $P_{\rm err}(k)$ leading to significant changes in its measurement. However, we caution that running of the bias parameters with scale by itself is not necessarily indicative of a breakdown in the bias expansion. The operators as defined in Eqn.~\ref{eqn:latetime} are correlated, and our procedure to estimate the bias parameters could be selecting a set $\hat{b}_i$ that runs with scale within a flat sub-region of the likelihood along the principal components of the Hessian, $M_{ij}$, with subsequent components that are poorly determined due to the the statistical uncertanties arising from a finite box volume. We report measurements carried out with subsets of the whole parameter set $\hat{b}_i$ and discuss the validity of second-order HEFT with constant bias parameters down to small scales in Appendices~\ref{appendix:A} and \ref{appendix:D}. The covariance of our bias parameter estimator $\hat{b}_i$ and correlations between bias parameters as a function of $\kmax$ are quantified and discussed in Appendix~\ref{appendix:C}.
\subsection{HODs, the halo model, and stochasticity}
\label{subsection:analyticnonpoiss}
When combined with the halo model of structure formation \citep{COORAY_2002}, HOD modelling provides analytic expressions for galaxy observables. The density contrast field is modelled as a mixture of the density of \emph{central} and \emph{satellite} galaxies, 
\begin{equation}
    \delta_g (\bk) = (1-f_{\rm sat}) \delta_c (\bk) + f_{\rm sat} \delta_s (\bk).
\end{equation}
The power spectrum of galaxies is thus decomposed into contributions that arise from central--central, central--satellite and satellite--satellite correlations 
\begin{align}
    P_{gg} (k) = (1 - f_{\rm sat})^2 P_{\rm cc} (k) + 2 f_{\rm sat}(1 - f_{\rm sat}) P_{\rm cs} (k)&\\ 
    \nonumber + f_{\rm sat}^2 P_{\rm ss}(k)&.
\end{align}
Each of these spectra, in turn, can be decomposed into contributions that arise from correlations in the \emph{one-halo} regime (that is, galaxies occupying the same halo) and the \emph{two-halo} regime (correlations between galaxies in separate halos). The one-halo contributions to each term can be written as
\begin{align}
    \label{eqn:pcc} &P_{\rm cc}^{{\rm (1h)}} = \frac{1}{\bar{n}_c} {}&\\
    \label{eqn:pcs} &P_{\rm cs}^{{\rm (1h)}} = \frac{1}{\bar{n}_c \bar{n}_s} \int dM n(M) \langle N_{\rm sat} \rangle (M) \langle N_{\rm cen} \rangle (M) \\
    \nonumber{}& \quad\quad\quad\quad\quad\quad \times  u (k | M) \theta(\langle N_{\rm sat} \rangle (M) - 1) \\
    \label{eqn:pss} &P_{\rm ss}^{{\rm (1h)}} = \frac{1}{\bar{n}_s} + \frac{1}{\bar{n}_s^2} \int dM n(M) \langle N_{\rm sat} \rangle (M) \left [ \langle N_{\rm sat} \rangle (M) - 1 \right ] \\
    \nonumber{}& \quad\quad\quad\quad\quad\quad  \times u^2 (k | M) \theta(\langle N_{\rm sat} \rangle (M) - 1),
\end{align}
where $n(M)$ is the halo mass function and $u(k|M)$ is the density profile of the halo. The specific functional form of the density profile does not matter for the purposes of investigating stochasticity on the scales considered in this work, however we assume that $\lim\limits_{k\to 0} u(k|M) = 1$ and $\lim\limits_{k\to \infty} u(k|M) = 0$. In the $k \to \infty$ limit the one-halo spectra predict the standard Poisson term
\begin{align}
    P_{gg}^{(1h)} (k) &\underset{k \to \infty}{=} \frac{(1-f_{\rm sat})^2}{\bar{n}_c} + \frac{f_{\rm sat}^2}{\bar{n}_s} \\
    &\underset{k \to \infty}{=} \frac{1}{\bar{n}_g}.
\end{align}
In the $k \to 0$ limit, the terms that depend on the halo density profile will contribute to the observed spectrum. In this limit, Eqns.~\ref{eqn:pcs},~\ref{eqn:pss} respectively contribute to the total power spectrum as 
\begin{align}
    \nonumber{}&2(1-f_{\rm sat})f_{\rm sat}P_{\rm cs}^{{\rm (1h)}}\underset{k \to 0}{=}  \frac{2}{\bar{n}_g^2} \int dM n(M) \langle N_{\rm sat} \rangle (M)  \\
    & \quad \quad\quad\quad\quad\quad\quad\quad\quad \times  \langle N_{\rm cen} \rangle (M)\theta(\langle N_{\rm sat} \rangle (M) - 1), \\
    \nonumber{}&f_{\rm sat}^2 P_{\rm ss}^{{\rm (1h)}} \underset{k \to 0}{=}  \frac{f_{\rm sat}}{\bar{n}_g} + \frac{1}{\bar{n}_g^2} \int dM n(M) \langle N_{\rm sat} \rangle (M)\\
    & \,\,\quad\quad\quad\quad  \times  \left [ \langle N_{\rm sat} \rangle (M) - 1 \right ] \theta(\langle N_{\rm sat} \rangle (M) - 1).
\end{align}
Including all terms together, the full $k\to 0 $ halo model + HOD spectrum is
\begin{align}
\label{eqn:k0superpoiss}
    P_{gg} (k)\nonumber{}&\underset{k \to 0}{=} \frac{1}{\bar{n}_g} + \frac{1}{\bar{n}_g^2} \int dM n(M) \langle N_{\rm sat} (M) \rangle  \\
    \nonumber{}&\, \times\theta(\langle N_{\rm sat} \rangle (M) - 1) \left [ \langle N_{\rm sat} (M) \rangle \right.  \\
    &\left. + 2 \langle N_{\rm cen} \rangle (M) - 1\right ].  
\end{align}
The second term is always positive, and can be thought of as being related to the variance of the satellite--satellite (or central--satellite) occupation relative to the average density. HODs that have most of their satellite occupation sourced from the high-mass tail of the mass function could then be expected to have this term comparable to the original Poisson prediction, sourcing a considerable amount of super-Poisson stochasticity at large scales. \par 
Eqn.~\ref{eqn:k0superpoiss} is particularly illuminating in the limit of a monochromatic mass function at a mass $M_h$, with the expected occupations satisfying $\langle N_{\rm cen} (M_h)\rangle = 1,\, \langle N_{\rm sat}(M_h) \rangle \geq 1:$
\begin{equation}
    n(M) = \bar{n}_h \delta^D (M - M_h),
\end{equation}
where $\delta^D(M - M_h)$ is a Dirac delta function at a fixed halo mass $M_h$ and $\bar{n}_h$ is the number density of haloes at this mass. The integral over the delta function results in the simplified expression
\begin{align}
    P_{gg} (k) & \underset{k \to 0}{=} \frac{1}{\bar{n}_g} + \frac{\bar{n}_h \langle N_{\rm sat} \rangle}{\bar{n}_g^2} \left [ \langle N_{\rm sat} \rangle (M_h) + 1 \right ] \nonumber \\
    \label{eqn:monochromsuperpoiss} &\underset{k \to 0}{=} \frac{1}{\bar{n}_g} \left ( 1 +  f_{\rm sat} \left [ \langle N_{\rm sat} \rangle (M_h) + 1 \right ] \right), 
\end{align}
where we used that $\bar{n}_h \langle N_{\rm sat} \rangle / \bar{n}_g = f_{\rm sat}$. Note that using $\bar{n}_g=\bar{n}_h(1+\langle N_{\rm sat} \rangle)$, Eqn.~\ref{eqn:monochromsuperpoiss} is equivalent to the shot-noise prediction from the number density of haloes $1/\bar{n}_h$. However, we choose to express it in the above form to make the connection to the HOD parameters clearer, as the equivalence is only formally true in the monochromatic limit where all haloes host a central galaxy. \par 
The above expression clarifies the analysis, for example, of \citet{Baldauf_2013} where it is noted that increasing the satellite fraction of an HOD can push the observed stochasticity to the super-Poisson regime. However, we see that it is not \emph{solely} the satellite fraction that controls this but instead the interplay between the steepness of the occupation and the satellite fraction that controls the amplitude of this super-Poisson contribution. That is, one could find samples with lower satellite fractions but larger deviations from Poisson shot noise than others, as we show is the case for the \redmagic sample described in \S~\ref{subsec:hodperr}.\par 
It is clear then that one can observe super-Poisson stochasticity in $P_{\rm err}$ even when all of the non-linear clustering contributions are taken into account in the model. This same one-halo term was previously noted to enhance the stochasticity of signals expected from line intensity mapping surveys \citep{Schaan_2021,dizgah2021precision}, and \cite{dizgah2021precision} have additionally explored how higher-order bias operators contribute to non-Poissonian noise in the context of line intensity mapping. \par
In the previous sub-section we also alluded to the importance of including the effect of \emph{halo exclusion} in the analysis of stochastic power spectra. In a sense, halo exclusion is the opposite effect to the previously discussed one-halo enhancement of stochasticity. Whereas the enhancement comes from multiple satellites contributing to self pairs at the same pixel, halo exclusion leads to a suppression of stochasticity due to the minimum distance scale imposed on halo correlations. In the simplified $k\to 0$ case with monochromatic mass function, the effect of exclusion is to decrease large-scale stochasticity. \par 
Following \cite{Baldauf_2013}, one can construct a toy model for exclusion by imposing a break in the correlation function at a radius $R_{\rm exc}:$
\begin{equation}
    \xi^{(d)}(r) = \begin{cases}
                    \xi^{(c)}(r) \, \text{ if }r \geq R_{\rm exc}, \\
                    -1  \, \text{ if }r < R_{\rm exc}, \\
                    \end{cases}
\end{equation}  
where $\xi^{(c)}(r)$ is the non-excluded two-point correlation function of the sample. The power spectrum under the presence of exclusion is then 
\begin{align}
    &P^{(d)}(k) = -\int_0^R d^3r j_0 (kr) + \int_R^\infty d^3r \xi^{(c)} (r) j_0(kr), \\
   \nonumber{} &P^{(d)}(k) = P^{(c)}(k)- V_{\rm exc} \biggl ( W_R(k) \biggr. \\
  \label{eqn:exclusion}&\biggl.\quad \quad\quad\quad+ \int \frac{d^3 q}{(2\pi)^3} P(k) W_R (|\bk - \bq|) \biggr),
\end{align}
where $j_0 (kr)$ zero-order spherical Bessel function, $W_R (k)$ is the Fourier transform of the top-hat window function and $V_{\rm exc}$ is the exclusion volume. For illustrative purposes here we take the exclusion volume to be $V_{\rm exc} = 4\pi R_h^3 / 3$ where $R_h$ is the spherical radius of a halo of mass $M_h$. In the $k\to 0$ limit  both the window function and convolution term contribute with an overall negative amplitude to the total signal. \par 
Adding both the enhancement and exclusion terms together we find
\begin{align}
\label{eqn:finalform}
P_{gg} (k) \underset{k \to 0}{=} &\frac{1}{\bar{n}_g} \left ( 1 +  f_{\rm sat} \left [ \langle N_{\rm sat} \rangle (M_h) + 1 \right ] \right)\\
\nonumber &- \frac{458}{1+\delta} \left ( \frac{M_h}{10^{13} {\rm M}_\odot h^{-1}} \right) \left ( \frac{\rm Mpc}{h} \right)^3, 
\end{align}
where $\delta$ represents the overdensity of a halo relative to the background density of the Universe, as we have re-written the exclusion volume in terms of the halo mass instead. While it is standard to take $1+\delta=200$ when treating haloes, \cite{Baldauf_2013} have shown that for exclusion radii at late times the exclusion term is better fit by assuming $\delta \approx 30$\footnote{The expression as presented in Eqn.~\ref{eqn:finalform} neglects the amplitude of the $k \to 0$ contribution from the convolution in Eqn.~\ref{eqn:exclusion}, and a lower $\delta$ would correspond to higher amplitude from this contribution.}. Nevertheless we may infer that for the case of an HOD, the properties of the galaxy sample control the super-Poissonianity, whereas the host haloes set the degree of sub-Poissonianity. This interplay between enhancement and exclusion will be crucial in interpreting the results of our subsequent numerical experiments.
\begin{figure*}
    \centering
    \includegraphics[width=0.8\textwidth]{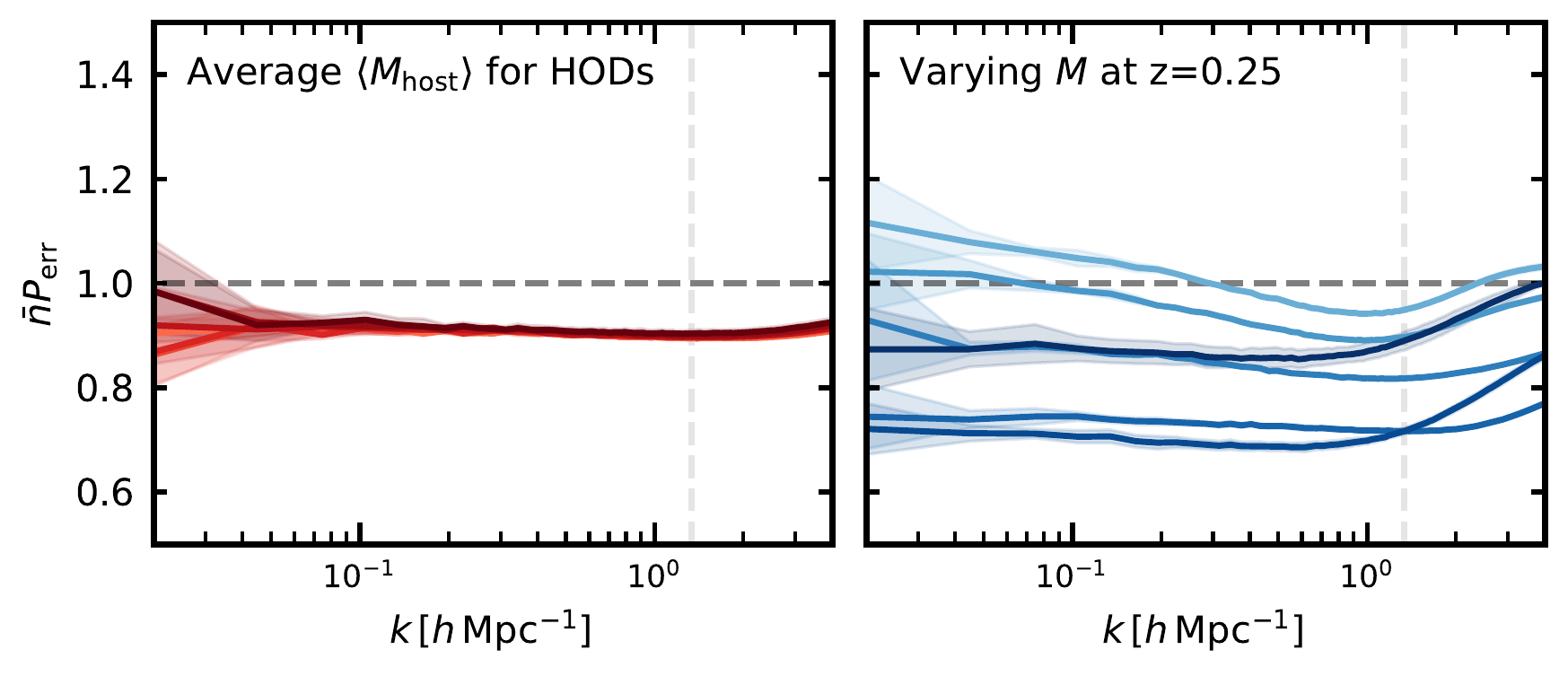}
    \caption{Error power spectra for two sets of halo samples. The power spectra are divided by the Poisson prediction, $\frac{1}{n}$, to highlight deviations from the standard expectation. The shaded regions correspond to one standard deviation from repeating this estimate for the 5 \aemulus  boxes that belong to our reference cosmology. The vertical dashed line corresponds to the inverse grid size, $L_{\rm cell}^{-1} \approx 1.33 \ihmpc$. \emph{Left panel:} halo mass bins, as described in Eqn.~\ref{eqn:mbins}, that encompass the average host halo mass for the HOD samples we consider in this paper in each snapshot, with the lightest (darkest) shade corresponding to the highest (lowest) redshift snapshots respectively. \emph{Right panel:} Varying halo mass in bins of 0.5 dex width for the snapshot at redshift $z=0.25$. The lightest (darkest) shade correspond to $\log M \in [12, 12.5]$ ($\log M \in [14.5, 15.0]$) respectively.}
    \label{fig:haloperr}
\end{figure*}
\section{Simulations and samples}
\label{sec:simsnsamples}
We use the \texttt{Aemulus} \citep{DeRose:2018xdj} suite of $N$-body simulations to populate three different samples of red galaxies using HODs. \texttt{Aemulus} is composed of 40+35 dark-matter-only $N$-body simulations with size $L_{\rm box} = 1050$ Mpc$~h^{-1}$ and $N=(1400)^3$ particles. They were designed to serve as a training set for the construction of emulators of non-linear summary statistics that could be readily applied to analysis of modern cosmological data sets. To date, these include emulators for the mass function \citep{McClintock:2018uyf}, halo bias \citep{mcclintock2019aemulus}, and the correlation function of HOD galaxies \citep{Zhai:2018plk}. We specifically choose a subset of the \aemulus suite that strikes a balance between being close to Planck's $\Lambda$CDM constraints but also offering repeated realizations so we may make statistical assessments of our results. Specifically, we pick the cosmology from the test suite that is closest to $\Lambda$CDM by relative differences. The parameters of this cosmology are $\Omega_c = 0.27$, $\Omega_b = 0.05$, $h = 0.6673$, $n_s=0.973$, $\sigma_8 = 0.798$, $N_{\rm eff} = 3.2$, $w=-0.9$.\par 
We use the public halo catalogs associated with these boxes and apply an HOD prescription to populate them with galaxies across the redshift range $z=[0.25, 1.0]$, which corresponds to most of the redshift range spanned by the samples of galaxies we wish to consider\footnote{This is done using \texttt{simplehod}, available at \href{https://github.com/bccp/simplehod}{https://github.com/bccp/simplehod}.}. We choose previously published HODs that describe three samples of red galaxies that are used for lensing and clustering analyses by current and next-generation surveys. These are: the \redmagic sample used in the Dark Energy Survey \citep{Rozo:2015mmv,Clampitt_2016, zacharegkas2021dark}; luminous red galaxies from BOSS and eBOSS \citep{Zhai_2017}; and DESI-like luminous red galaxies from \citet{Zhou_2020}. The parametric forms adopted are all variations of the standard \citet{Zheng:2004id,Zheng:2007zg} HOD discussed in \S~\ref{subsec:HOD} and displayed in Eqns.~\ref{eqn:ncen} and \ref{eqn:nsat}. For all of our samples in question, we note that their auto-power spectra become dominated by shot noise at similar scales, at around $k\sim 0.4 \ihmpc$. Since we use field-level information and not just $P_{gg}$ in our analysis, we believe this justifies our fiducial choice of $\kmax$ in the analysis below.\par 
The samples we mock up are representative of those observed by leading galaxy surveys.  While the HOD parameters adopted are not calibrated to the specific cosmology we use, the impact of this difference is sub-leading for the kind of analysis we wish to carry out for the BOSS and DESI samples, as their derived parameters we obtain are statistically consistent with those reported in the original publications. For the case of the DES Y3 HOD presented in \cite{zacharegkas2021dark}, when applied to our fiducial cosmology, we find significantly different satellite fractions, mean halo masses, and number densities compared to the published results. \par 
The differences between these results and ours arise from differences in the high-mass tail of the halo mass function of our simulations compared to the \cite{Tinker:2008ff} mass function adopted in the original publication at their fiducial cosmology. The large satellite slopes ($\alpha > 1.6$ for all bins) and suppression of occupation until high masses leads to significant differences in derived parameters for the galaxy samples in our simulations. \par 
To alleviate these discrepancies, we have slightly tuned the parameters of the DESY3 \redmagic HOD in a way that recovers the satellite fractions, mean host halo masses, and number densities reported in \cite{zacharegkas2021dark} in our simulations. We show the resulting HOD compared to the original in Fig.~\ref{fig:desappendix}. The largest change is a reduction in $\alpha$ across all redshift bins, as well as a slight boost to the halo occupations at lower masses. \par 
We interpret our results on galaxy stochasticity by analyzing the dependence of  stochastic power spectra on three derived HOD parameters: satellite fraction $f_{\rm sat}$, galaxy number density $\bar{n}$ and average host halo mass $\log_{10} \langle M_{\rm host} \rangle$. The results we find at each redshift are reported in Tables~\ref{table:redmagic}, \ref{table:boss} and \ref{table:desi}. The corresponding galaxy catalogs span an order of magnitude in density and halo mass, with $\bar{n} \in [1.39, 9.75] \times 10^{-4} \, [h {\rm Mpc}^{-1}]^3$ and $\log_{10} \langle M_{\rm host} / (h^{-1} M_\odot) \rangle \in [12.95, 13.64]$. We plot the galaxy occupations of our HOD samples for each snapshot in Fig.~\ref{fig:hodplots}. This figure highlights a few key differences in how our mock galaxies populate their haloes, which we use to interpret our findings on stochasticity. Most notably, the Y3-like \redmagic HOD has significantly lower occupations than the other samples until higher masses. The \redmagic galaxies have high number density, and this deficiency at low mass is made up by having a larger slope in the satellite occupation. This is noticeable in the bottom three panels of Fig.~\ref{fig:hodplots}. On the other hand, the BOSS and DESI HODs share similar derived parameters with the largest difference being the significantly higher number density of DESI galaxies.\par 
Both the DESI-like LRGs and \redmagic galaxies have constrained the redshift-dependence of the fundamental HOD parameters. For the BOSS sample, we fix its parameters as a function of redshift. However, as the DESI LRG sample infers little redshift dependence in its own parameters we believe this is a suitable approximation for the intent of our analysis. \par
\begin{figure*}
    \centering
    \includegraphics[width=\textwidth]{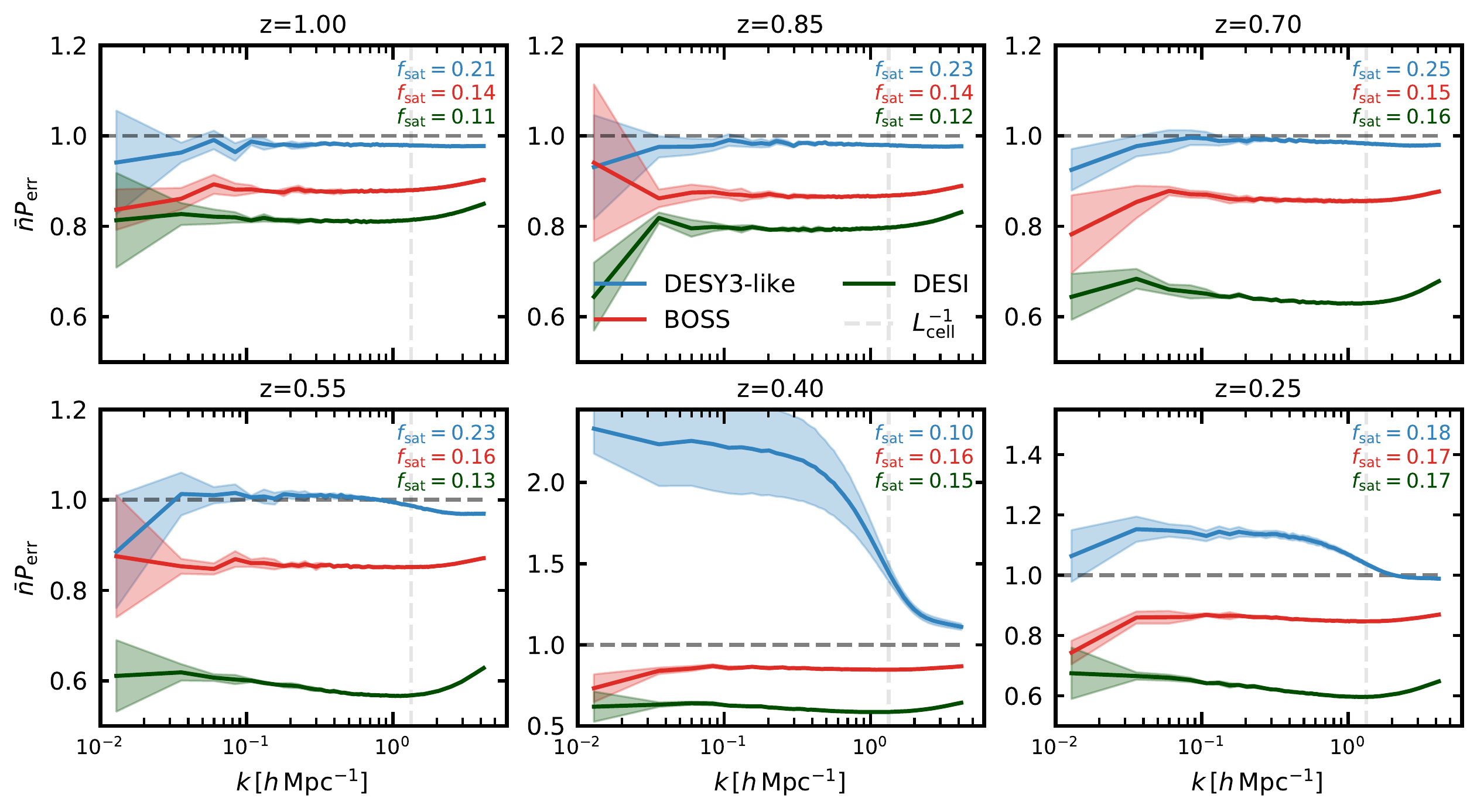}
    \caption{Error power spectra for the different samples, for the standard Lagrangian bias basis, using bias parameters inferred from our variance-minimization procedure assuming $k_{\rm max} = 0.4 \ihmpc $. The envelopes are the scatter found from the five independent realizations from the Aemulus suite at this cosmology.  The vertical dashed line corresponds to the inverse grid size, $L_{\rm cell}^{-1} \approx 1.33 \ihmpc$. Note that at low redshifts the $y$-axis ranges are altered to accommodate for the large amount of super-Poisson stochasticity observed in the \redmagic sample.}
    \label{fig:hod_perr}
\end{figure*}
\section{Results}
\label{sec:results}
\subsection{Sub-poisson stochasticity of massive halos}
\label{subsec:haloperr}
The stochastic power spectrum of dark matter haloes has been the subject of considerable past study \citep{Hamaus_2010,Baldauf_2013}. Notably, \cite{Schmittfull_2019} measured these statistics for halo samples that collectively spanned four orders of magnitude in halo mass. They observed that less massive haloes tend to
exhibit super-Poisson stochasticity\footnote{We define super and sub-Poisson stochasticity as the regimes where $\bar{n}P_{\rm err} > 1$ and $\bar{n}P_{\rm err} < 1$ respectively.}, which trends toward sub-Poisson with increasing mass. These results were obtained using a different model for the field-level tracer density, and so in this section we report our results for the halo samples that host the galaxies using the second-order HEFT model that is the subject of study of this paper. While \citet{Schmittfull_2019} used a third-order Eulerian bias model, we work with the second-order Lagrangian bias expansion of Eqn.~\ref{eqn:secondorder}, which has been shown in the past to capture the clustering statistics of halos in the mass ranges under consideration \citep{Abidi:2018eyd}. Another key difference is that we use the fully non-linear Lagrangian displacement field as determined from the $N$-body simulation, whereas \cite{Schmittfull_2019} only includes Zel'dovich displacements in their shifted operators. \par
From Tables~\ref{table:redmagic}, \ref{table:boss}, \ref{table:desi} we can infer that for any given snapshot the average host halo mass spans approximately 0.2 dex for all of the HOD samples we have constructed. At a given snapshot we select for halos in the mass bin given by 
\begin{equation}
\label{eqn:mbins}
13.1 + \frac{2}{3}(1-z) \leq \log M < 13.3 + \frac{2}{3}(1-z),
\end{equation}
where $z$ is the redshift of the snapshot. These mass ranges encompass the average halo masses of our HOD samples. \par 
We proceed to measure $P_{\rm err}(k)$ for these halo samples using the procedure outlined in \S~\ref{subsec:Perrstimate}, and report our results in the left panel of Fig.~\ref{fig:haloperr}. For all snapshots under consideration we observe a slight but significant sub-Poisson signal, in concordance with previous results in the literature for comparable halo masses. For our second-order basis, the amplitude of the deviation from the Poisson expectation is approximately 10 per-cent, staying approximately constant with redshift. Qualitatively, this can be understood by noting that we simultaneously increase the mean halo mass across snapshots, which increases the expected exclusion signal, while simultaneously going down the mass function to less dense halo samples, which increases the one-halo enhancement expected. \par 
We also observe the $P_{\rm err} (k)$ we have measured are approximately scale independent out to scales comparable to the inverse of the grid spacing, at which point we observe an uptick due to the impact of deconvolving with the window function used to assign objects to the grid. The scale independence of the $P_{\rm err}(k)$ indicate second-order HEFT is a suitable forward model for the halo sample under consideration. These initial results highlight consistency between hybrid EFT approaches to estimating $P_{\rm err}(k)$ and others in the literature, such as the shifted-Eulerian operator basis of \cite{Schmittfull_2019}. \par
As an additional test of second-order HEFT, we also look at the evolution of halo stochasticity as a function of mass, at fixed redshift. Specifically, we select six broad bins in halo mass, from $10^{12} h^{-1}M_\odot$ to $10^{15} h^{-1}M_\odot $ with width of 0.5 dex. We measure their stochastic power spectra and show them in the right panel of Fig.~\ref{fig:haloperr}. The purpose of these measurements are twofold: they help us verify the validity of the approximate exclusion treatment as described in \S~\ref{subsection:analyticnonpoiss} and also how well can second-order HEFT describe different halo samples of varying mass bins. If halo exclusion scales roughly linearly with halo mass, as shown in Eqn.~\ref{eqn:finalform}, then in the power-law regime of the mass function, the exclusion signal should get larger with mass bin. However, once the mass bin reaches the exponential cut-off of the mass function,  the number density should fall off faster than $M$ and the stochastic power spectrum should begin to revert to Poisson. This is precisely the behavior we observe, where the line corresponding to the highest mass bin at $M \in [10^{14.5}, 10^{15}]\, h^{-1} M_\odot$ has its stochasticity closer to Poisson than the previous bin. \par 
We also note that the lowest mass bins in the right panel of Fig.~\ref{fig:haloperr} show a slight tendency toward super-Poissonianity but also with significant scale dependence. These results ostensibly show that lower-mass haloes are potentially more sensitive to neglected higher-order bias contributions that manifest themselves as super-Poissonianities atop the exclusion signal. However, a quantitative assessment of the impact of higher-order operators is postponed to future work, because for the halo masses relevant to our HODs the stochastic power spectra of host haloes are relatively scale independent.
\subsection{The stochasticity of red survey galaxies}
\label{subsec:hodperr}
Having established a baseline for the degree of deviation from Poisson stochasticity in our host halo samples, we can now turn to analyzing the mock galaxy samples we have created. In Fig.~\ref{fig:hod_perr} we report the redshift-dependent product $\bar{n}(z) P_{\rm err}(k,z)$ for the three red galaxy samples we consider, again obtained using the procedure outlined in \S~\ref{subsec:Perrstimate}. If the impact of enhancement and exclusion on these galaxies were negligible, we would expect $\bar{n} P_{\rm err}(k)=1$. \par 
We find that for most of the above galaxy samples, the stochastic power spectra are approximately scale independent across the range of scales considered. Even though all of our mock galaxies populate comparable dark matter haloes, whose stochasticity is significantly sub-Poisson, we in fact observe that the three HOD parameterizations adopted exhibit markedly different levels of stochasticity.  Once again, this points to the suitability of second-order Lagrangian bias in describing these samples\footnote{ Additionally, we note that for the case of \redmagic galaxies at $z=0.55$ the inferred bias parameters from the field-level fitting procedure are consistent with those obtained by fitting galaxy--galaxy and galaxy--matter spectra in \cite{Kokron_2021}, providing additional validation of our methodology.}. At high redshifts, the \redmagic sample exhibits a slight sub-Poisson trend but over time has $\bar{n} P_{\rm err}$ evolve to strongly super-Poisson, with a strong scale-dependence that reverts to close to the Poisson measure at small scales. This result can be explained by the analysis carried out in \S~\ref{subsection:analyticnonpoiss}, combined with knowledge of the HOD of the Y3 \redmagic sample. From Eqn.~\ref{eqn:monochromsuperpoiss}, the steeper the satellite occupation at a given mass, the larger its contribution is to the one-halo enhancement that drives the stochasticity to be super-Poisson. From Fig.~\ref{fig:hodplots} and Table~\ref{table:redmagic} the galaxy samples at $z=0.4$ and $z=0.25$ have both simultaneously suppressed occupation until high masses, and then significantly larger slopes than the other samples to try and fit to the observed number density. The panels that exhibit this super-Poisson stochasticity have the highest slopes relative to the other bins, and the reversion to close to $nP \sim 1$ is indicative of the halo density profile decaying at small scales as expected. Thus, we can conclude that the lower-redshift \redmagic samples exhibit a super-Poisson signal. \par
In fact, this super-Poissonian aspect of \redmagic stochasticity has been previously observed in the literature \citep{Friedrich_2018, Gruen_2018, friedrich2021pdf}. We also note the snapshot at $z=0.40$, corresponding to the second \redmagic lens bin, is anomalous in its behavior. The degree of super-Poisson stochasticity is markedly higher, and the measurements themselves are quite noisy. The second lens bin in DES Y3 has by far the highest slope $\alpha$ in its satellite distribution. This means that a small number of very massive haloes host satellites. The number of these high mass haloes varies significantly between our five realizations, which  explains the increased scatter we observe only for this snapshot for the \redmagic HOD.\par 
Turning to the sample of BOSS-like LRGs, we observe mild disagreement with the Poisson prediction across a wide range of scales for all of the snapshots under consideration. The snapshots possess a slightly sub-Poisson stochastic power spectrum, however to a mild degree when compared to the other two samples -- on the order of 10 per-cent. In this case, numerical estimates of the amplitude of one-halo enhancement for BOSS galaxies through Eqn.~\ref{eqn:k0superpoiss} show that it is entirely negligible, as at no masses in our halo catalog does this HOD have $\langle N_{\rm sat} \rangle \geq 1$. Given that the satellite occupation is sub-leading at all masses, this is not surprising. The mild sub-Poissonianity we observe corresponds solely to the imprint of halo exclusion on this sample. \par 
The DESI LRGs, on the other hand, exhibit significant sub-Poisson stochasticity. This is despite the fact that their derived parameters are quite similar to that of BOSS; both satellite fractions and host halo masses are very comparable on a per-snapshot basis. In fact, the occupations for both samples are quite similar except for the fact that the DESI occupation is almost multiplicatively offset from the BOSS one. If both samples are hosted within similar haloes, then, we expect their exclusion signatures to be comparable. However, as the DESI galaxies are significantly denser, the relative size of the exclusion signal compared to the galaxy density is larger (see Eqn.~\ref{eqn:finalform}) and we observe a more sub-Poisson stochasticity as a result. \par 
The significantly higher number density of the DESI HOD could imply additional bias terms not included in our model that could contaminate our estimate of $P_{\rm err}(k)$. This, then, would explain the stronger deviations from a flat noise curve at large scales for the DESI sample relative to other samples.  However, as third-order bias models introduce a significant number of new parameters and our simulations are of limited volume, we defer the investigation of the impact of including higher-order HEFT operators to future work. \par 
\subsection{What range of deviations from Poisson stochasticity is allowed when conditioned on an HOD?}
\label{subsec:chainhod}
\begin{figure}
    \centering
    \includegraphics[width=\columnwidth]{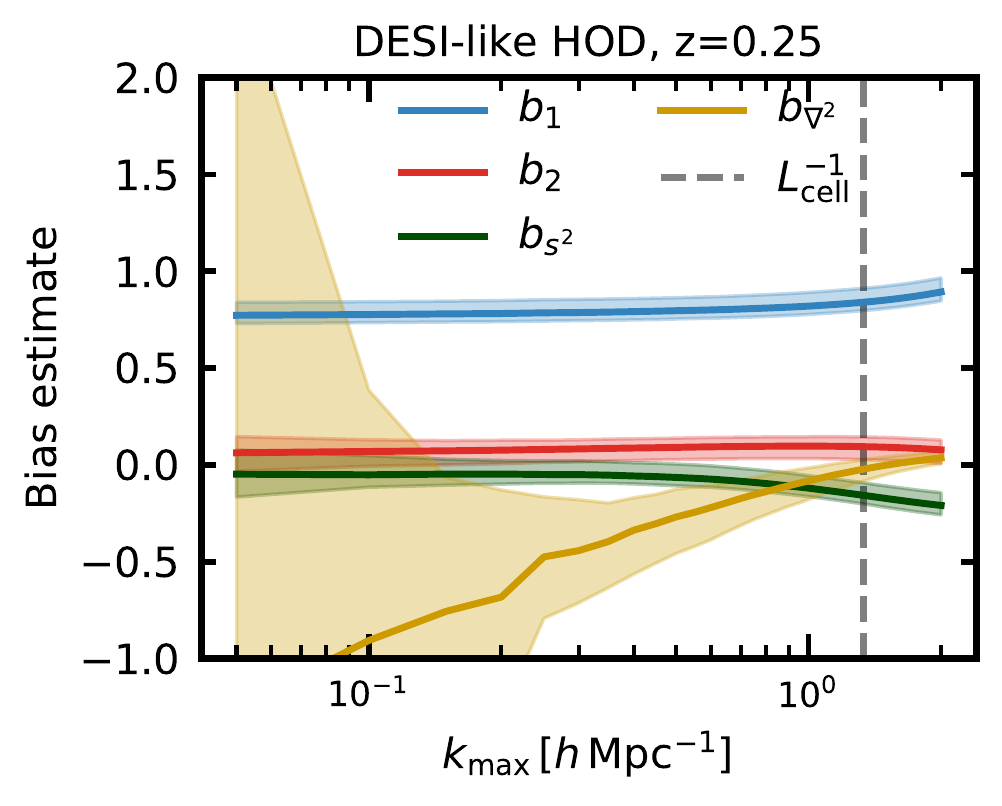}
    \caption{Estimates of second-order Lagrangian bias parameters obtained from $N=1500$ mock galaxy samples from the DESI-like LRG posteriors, as a function of maximum wavenumber $k_{\rm max}$ used in the fit. The solid line shows the median bias parameter and the shaded regions the 98\% quantiles. The vertical dashed line corresponds to the inverse grid size, $L_{\rm cell}^{-1} \approx 1.33 \ihmpc$.}
    \label{fig:hodchainbias} 
\end{figure}
The numerical experiments carried out in the previous sub-section are concerned with the expected degree of stochasticity subject to a specific set of HOD parameters. In principle, one would expect that degeneracies in the HOD lead to a population of HOD parameters that can equally capture the statistical properties of a galaxy sample. However, for any given set of parameters we will find a value of stochasticity, and thus we should also expect that as a result any given sample is consistent with a range of large-scale stochastic behavior. \par 
To assess this, we take a deeper look at the set of parameters consistent with the DESI-like HOD of \cite{Zhou_2020} for their lowest redshift bin. We take 300 random samples from the post burn-in MCMC chains run in the publication and use the procedure laid out in \S~\ref{subsec:Perrstimate} to estimate the low-$k$ stochasticity for that point in the chain. For each sample, we populate galaxies across all five boxes at our fiducial cosmology, leading to a total of $N=1500$ sets of bias parameters and measurements of $\bar{n}P_{\rm err}$ that to some extent also includes a contribution from cosmic variance. \par 
In Fig.~\ref{fig:hodchainbias} we report the estimated bias parameters that we obtain from this procedure. Interestingly, the first three parameters $b_1, \, b_2, \, b_{s^2}$ are approximately constant out to very small scales, beyond the fiducial $k_{\rm max} = 0.4 \ihmpc$ value we use, and their 98\% quantiles show very little spread in the estimates. For $b_{\nabla^2}$ we find a significantly larger scatter, that decreased as we increase $k_{\rm max}$, and a median value that also seems to run more steeply with scale than other parameters. The larger scatter in $b_{\nabla^2}$ seems to indicate this field possesses a larger correlation between realization-to-realization noise. This is somewhat expected given the origin of this operator, as it arises from integrating out the untameable effects of small scales. \par
\begin{figure}
    \centering
    \includegraphics[width=\columnwidth]{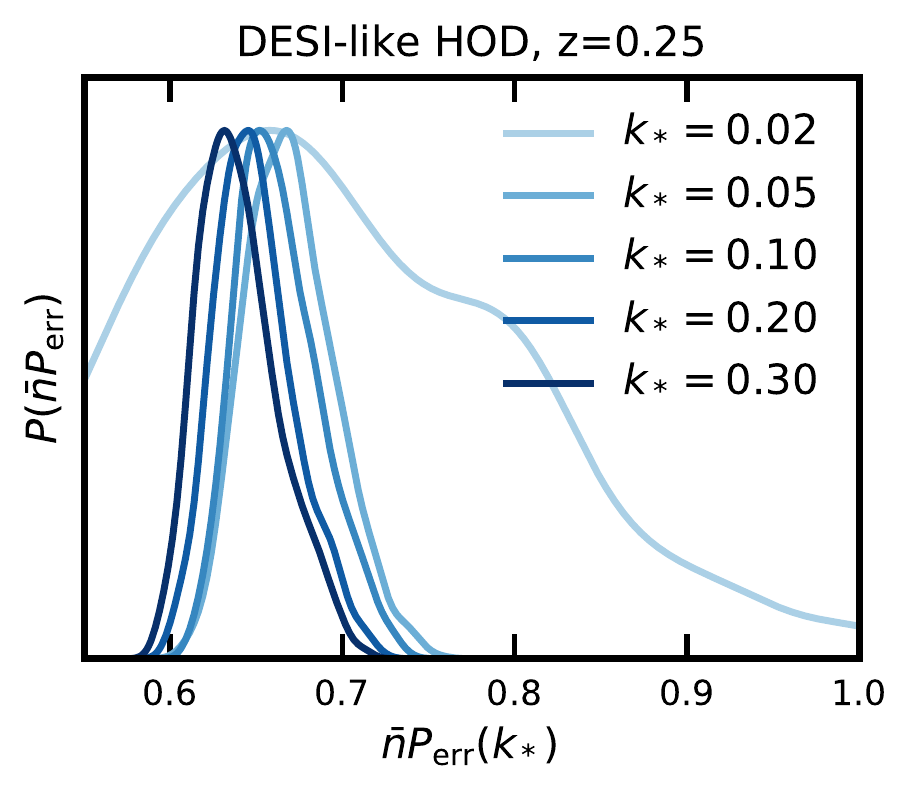}
    \caption{Distributions of stochastic power spectra $\bar{n}P_{\rm err}(k)$ obtained from $N=1500$ mock galaxy samples from the DESI-like LRG posteriors. Each curve represents a different slice of the stochastic spectrum evaluated at a different wavenumber, in units of $\ihmpc$.}
    \label{fig:nperrdistrib} 
\end{figure}

 
As an additional test, we investigate the posterior distribution of stochasticities consistent with DESI-like LRGs, $P(\bar{n}P_{\rm err})$ sliced through different scales $k_*$. All of the $\bar{n}P_{\rm err}$ are measured with bias parameters estimated at $k_{\rm max} = 0.4 \ihmpc$, our fiducial choice for this publication. The resulting distributions are shown in Fig.~\ref{fig:nperrdistrib}. Slicing through the stochastic spectra at large scales shows a wide distribution of values. However, this is not surprising as our boxes have limited volume and scales such as $k \sim 0.02 \ihmpc$ will still be affected by cosmic variance. As we probe the distribution toward quasi-linear scales we see that the allowed range of stochasticity remains relatively peaked around a single value, while the value of the peak itself shifts very slightly. \par
Thus, we can expect that the range of HODs consistent with a given data set will have relatively similar large-scale stochasticities, more tightly spread than what is observed as we look across HODs that describe different galaxy samples. 
\subsection{Priors on red galaxy stochasticity}
\label{subsec:priors}
The results presented in this section highlight that analyses of red galaxy samples using second-order Lagrangian bias models can safely control the impact of stochastic contributions to model power spectra. The detections of deviations from Poisson stochasticity are significant, and their amplitudes span the range of being 60--140 per-cent of the Poisson expectation, with the exception of the anomalous $z=0.4$ \redmagic sample. Cosmological parameter inference carried out with comparable models and samples of galaxies can adopt informative priors on the degree of stochasticity given certain knowledge about the selection of the sample. Priors can be adopted as being of the form
\begin{equation}
    P(a_1) = \mathcal{N} \left(1, 0.4 \right ), 
\end{equation}
where $a_1$ is the scale-independent stochastic amplitude from Eqn.~\ref{eqn:poisson}. This prior encompasses the whole range of deviations from Poisson stochasticity observed across the redshift range $0.25 \leq z \leq 1$ within $1-\sigma$, with the exception of the $z=0.40$ bin of the \redmagic sample, which possesses an anomalously high super-Poisson stochasticity. If the sample is constrained to higher redshifts $z>0.55$ then setting $\sigma_{a_1} \approx 0.3$ gives a tighter prior that still captures the observed deviations from Poisson stochasticity to within $1-\sigma$. 
\begin{table}
\centering
\begin{tabular}{c c c c} 
 \hline \hline
 z & $f_{\rm sat}$ &  $10^4\bar{n}$ & $ \log_{10}\langle M_{\rm host} \rangle$\\ 
1.00 & 0.21  &  3.66  &  12.95  \\  
0.85 &0.23   & 3.99  &  13.06    \\
0.70 &0.25  &  4.33  &  13.18    \\
0.55 &0.23  &  9.70  &  13.24    \\
0.40 &0.10  &  9.22  &  13.51    \\
0.25 & 0.18   & 9.75 &   13.51    \\
 \hline \hline
\end{tabular}
\caption{Derived parameters for the \redmagic HOD.}
\label{table:redmagic}
\end{table}

\begin{table}
\centering
\begin{tabular}{c  c c c} 
 \hline \hline

 z & $f_{\rm sat}$ & $10^4\bar{n}$ & $ \log_{10}\langle M_{\rm host} \rangle$\\
1.00 &0.14  &  1.39  &  13.32    \\
0.85 &0.14  &  1.62   & 13.37    \\
0.70 &0.15  &  1.88    &13.43    \\
0.55 &0.16  &  2.15   & 13.49    \\
0.40 &0.16  &  2.44   & 13.55  \\  
0.25 &0.17 &   2.74   & 13.61\\    
 \hline \hline
\end{tabular}
\caption{Derived parameters for the \texttt{BOSS} LRG HOD.}
\label{table:boss}
\end{table}
\begin{table}
\centering
\begin{tabular}{c c c c} 
 \hline \hline 
 z & $f_{\rm sat}$ & $10^4\bar{n}$ & $ \log_{10}\langle M_{\rm host} \rangle$\\ 
1.00 &0.11  &  2.17 &   13.29    \\
0.85 &0.12  &  2.53  &  13.36    \\
0.70 &0.16  &  6.69  &  13.39    \\
0.55 &0.13  &  5.88  &  13.50   \\ 
0.40 &0.15 &   7.57  &  13.55 \\   
0.25 &0.17&   8.50 &  13.64\\   
\hline \hline
\end{tabular}
\caption{Derived parameters for the \texttt{DESI} LRG HOD.}
\label{table:desi}
\end{table}

\section{Conclusions}
\label{sec:conclusion}
In this work we have applied a field-level model for biased tracers to study the stochastic contribution to the tracer--matter connection. Specifically, we focused on the stochasticity of both halo samples and example samples of red galaxies that are hosted in these halos, using three different forms of halo occupation distributions. The HODs we adopted have been previously used to describe the clustering statistics of bright red galaxy samples from three different galaxy surveys. \par 
We use second-order Hybrid Effective Field Theory to model the distribution of these galaxies at the field level. We developed an estimator that can reliably infer the maximum-likelihood bias parameters from the field-level information and applied it to infer the stochastic power spectra of these galaxy samples. We proceeded to compare our results to the commonly used assumption of Poisson stochasticity, framing our findings within the context of the well-established halo model of large-scale structure. Our findings can be summarized as follows:
\begin{enumerate}
    \item Almost all of the HODs used deviate from the constant, Poisson prediction of shot-noise by at most 40\%. This implies tight priors can be used on the expected stochasticity from these kinds of samples. This will reduce degeneracies between stochasticity, bias parameters, and cosmological parameters. 
    \item The form of the HOD is connected to the degree of non-Poisson stochasticity. Specifically, HODs with a high variance in their satellite occupation, or equivalently large slopes, will have super-Poisson stochasticity due to one-halo enhancements to the signal. 
    \item Very dense galaxy samples with negligible one-halo enhancements will instead have their large-scale stochasticity dominated by the effects of halo exclusion. As a result they will tend to have sub-Poisson stochasticity.
    \item The stochastic power spectra of galaxy samples can be either sub- or super-Poisson, despite being hosted in similar halo populations whose own stochasticities are consistently sub-Poisson. 
\end{enumerate}
These findings showcase the synergistic gains to be obtained from jointly studying models of the galaxy--halo connection with different models for modeling large-scale structure and bias. The combination of an empirical parameterization with a Lagrangian bias model allowed us to quantify the degree of stochasticity of these galaxies and place informative priors on this that will help future analyses of galaxy survey data. \par 
The 40\% priors highlighted in this publication highlight the wealth of stochasticities that red galaxies samples can exhibit. They are directly related to the fundamental scatter in allowed stochasticities of the physical process underlying galaxy formation. This implies narrowing the priors without making further assumptions about the sample will be challenging. Efforts to tighten these priors would require going beyond our analysis and carrying out more systematic studies for an intended sample. For example, the analysis of \S~\ref{subsec:chainhod}, where we found that conditioned on HODs consistent with DESI-like red galaxies, the spread of stochasticities was significantly tighter than 40\%. \par
The techniques to estimate field-level residual maps of tracer stochasticity developed here can be extended further. While we have limited ourselves to studying the auto-spectra of our residual maps, there are several avenues of investigation that can be pursued and are highly relevant to modern galaxy surveys. For example, one could characterize the cross-spectra of stochasticity between different tracer samples such as galaxies and clusters, whose cross-correlations are a promising future probe of cosmology \citep{To_2021} but whose cross-stochasticity is poorly understood. As analyses of higher $N$-point correlation functions become more prominent, the residual maps created here could shed new insights into the significantly more complicated case of stochasticity in higher $N$-point functions as well as be used to better understand when a second-order hybrid EFT model is no longer sufficient in describing a sample. \par 
This study also points to the feasibility of a broader program of learning about the galaxy--halo connection using bias models and empirical models in tandem, at the field level, to gain insights into how tracers relate to the distribution of dark matter at large. The analysis carried out here can be extended to different forms of tracer samples. For example, one could study the relationship between assembly bias and the Lagrangian bias parameters within HEFT. First steps in this direction, albeit for a different forward model, were carried out in \citet{lazeyras2021assembly}. HEFT could also be used to study HODs of different samples of galaxies that are not captured by the standard \citet{Zheng:2004id} form, and place priors on the bias parameters expected in that case. Similar work, for galaxies in \texttt{IllustrisTNG}~\citep{nelson2021illustristng}, has been explored recently \citep{barreira2021galaxy}. As this work was being finalized, \cite{zennaro2021priors} used hybrid EFT models precisely in this way to study the distributions of second-order Lagrangian bias for samples of galaxies populated using an extended sub-halo abundance matching scheme. Connecting efforts to measure bias parameters within different bias models to each other is another important goal that will aid in characterizing the relationship between empirical models, bias models, and their applicability to optimizing analyses of data collected by galaxy surveys.

\section*{Acknowledgements}
We would like to thank Rongpu Zhou for making the DESI-like LRG posteriors available, Georgios Zacharegkas for helpful discussions about the Y3 \redmagic HOD, and Sihan Yuan for comments on a draft of this work. We are grateful to the \texttt{Aemulus} collaboration for making the simulation suite used here publicly available. This work was supported in
part by U.S. Department of Energy contracts to SLAC (DE-AC02-
76SF00515). NK acknowledges support from the Gerald J. Lieberman fellowship. SC and MW acknowledge support from the National Science Foundation and Department of Energy. JD is supported by the Chamberlain fellowship at Lawrence Berkeley National Laboratory.
This research has made use of NASA's Astrophysics Data System and the arXiv preprint server. \par
Some of the computing for this project was performed on the Sherlock cluster at Stanford. We would like to thank Stanford University and the Stanford Research Computing Center for providing computational resources and support that contributed to these research results.\par
Calculations and figures in this work have been made using \texttt{nbodykit} \citep{Hand_2018}, \texttt{GetDist} \citep{lewis2019getdist}, and the SciPy Stack \citep{2020NumPy-Array,2020SciPy-NMeth,4160265}.  
\section*{Data Availability}
The data underlying this article are available in the \href{https://aemulusproject.github.io/}{ \texttt{Aemulus} Project's} website. Access to the HEFT products is available on reasonable request.
\bibliography{main}
\bibliographystyle{mnras}

\appendix

\section{The scale-dependence of bias parameters}
\label{appendix:A}
\begin{figure}
    \centering
    \includegraphics[width=1.\columnwidth]{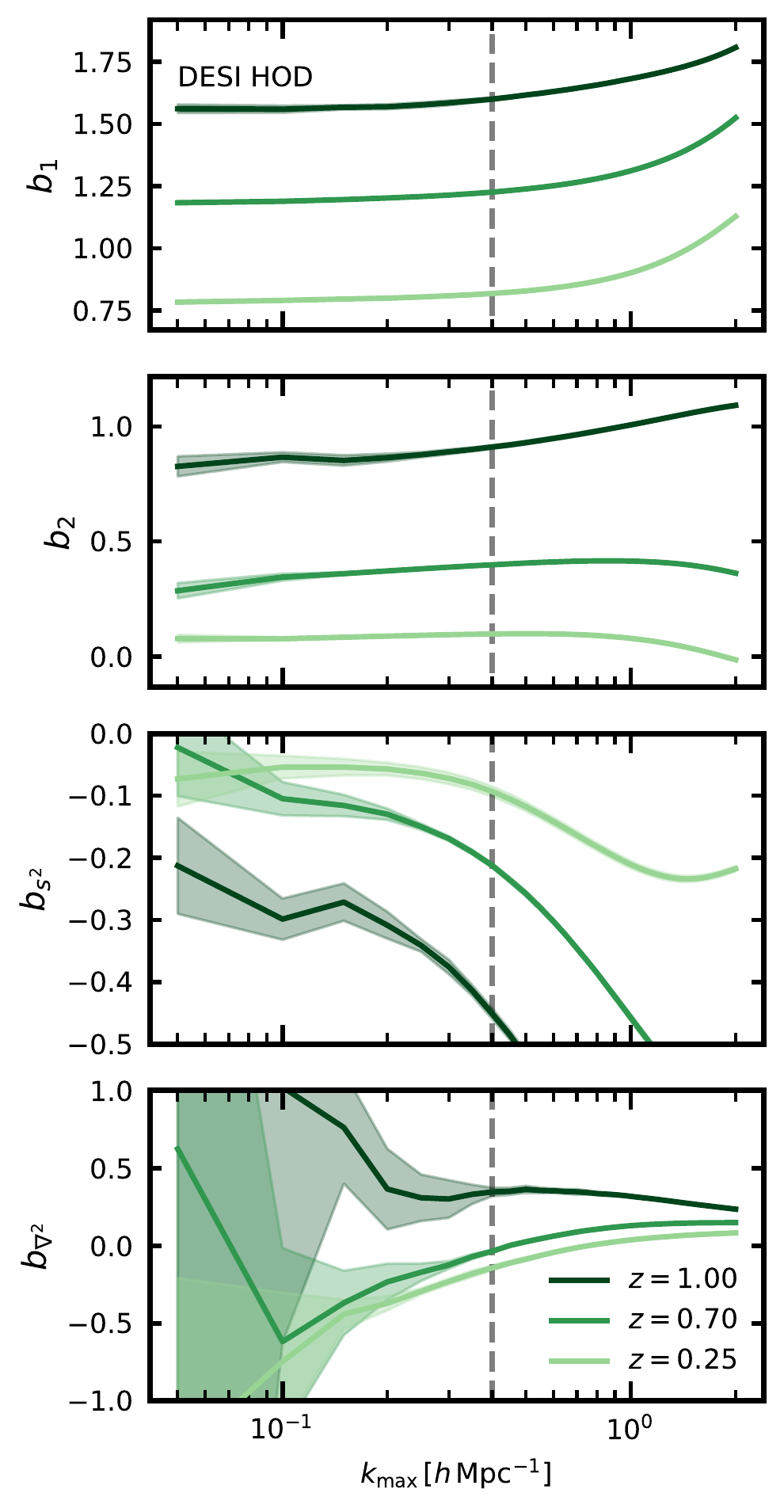}
    \caption{Scale-dependent estimates of Lagrangian bias parameters for the DESI HOD, for three snapshots used in this analysis. The figure shows both the regimes over which we can trust the bias model and also the redshift-dependence of the estimated bias parameters.}
    \label{fig:biastransfer}
\end{figure}

In this appendix we report our results on the estimates of the bias parameters $\hat{b}_i$. For each snapshot and HOD form we estimate the $\hat{b}_i$ following Eqn.~\ref{eqn:bhat}. The assumption that second-order HEFT describes the galaxy sample well can then be tied to the range of scales at which the coefficients remain scale independent. We show these results in Fig.~\ref{fig:biastransfer}. We show the results for the DESI sample at $z=[1.0, 0.7, 0.25]$. The results are broadly consistent with the expectations from biasing theory -- the amplitudes of the bias coefficients decrease as we arrive at later times and consequently less-biased samples. We also note that with the exception of the quadratic bias parameter $b_{\nabla^2}$, even using $\kmax = 0.1$ leads to highly precise determinations of the bias parameters despite the limited volume of our boxes. This is due to the field-level nature of the fit, which includes significantly more cosmological information than simply finding bias parameters by fitting summary statistics such as the clustering and galaxy--matter power spectra.\par 
We also find, again with the exception of $b_{\nabla^2}$, that the inferred parameters remain relatively stable out to very high $\kmax$ across all of our redshift bins. We also find some slight running for the tidal bias $b_{s^2}$ at $k \gtrsim 0.4 \ihmpc$ for some bins, which motivates our fiducial choice of $\kmax = 0.4 \ihmpc$ in this publication. However, we note that the procedure selects a single set of biases that minimizes $P_{\rm err}(k)$. As the operators are correlated, there could exist a separate set of $\hat{b}'_i$ that are scale independent and result in an equally acceptable $P_{\rm err}(k)$. Indeed, the fact that at $k_* = 0.3 \ihmpc$ and $k_* = 0.5 \ihmpc$ we find statistically indistinguishable stochastic power spectra supports this argument. \par 
The results of Fig.~\ref{fig:biastransfer} indicate that our estimation procedure for bias parameters from field-level data is both precise and robust. While in the main publication we concern ourselves mainly with $\bar{n} P_{\rm err}$ there is considerable interest in applying these techniques to study the actual parameters for a varying class of tracer--matter connection models. \par 
We also note that the fact that the bias parameters begin to run at $\kmax \sim 0.4 \ihmpc$ is not inconsistent with the results of the work of \cite{Kokron_2021} which fit power spectra to $\kmax = 0.6 \ihmpc$ (and recently \cite{zennaro2021priors} which go to even smaller scales). Those publications were concerned with fitting the clustering and lensing power spectra, while in this publication we concern ourselves with the significantly more difficult problem of describing the full statistical properties of the field that encapsulates information from \emph{all} $N$-point functions. Indeed, similar tests were carried out in \cite{banerjee2021modeling} with statistics that also encode higher $N$-point information and HEFT was found to be a good fit to them at slightly more conservative scales than what was found for power spectra. 

\section{The dependence of $\bar{n}P_{\rm err}$ on $k_{\rm max}$}
\label{appendix:B}
In our procedure to obtain the error power spectrum $P_{\rm err}(k)$ we must choose a cut-off scale $k_{\rm max}$ that indicates the smallest scales used to estimate $\hat{b}_i$ and construct a field-level realization of that galaxy sample, under the assumption of \emph{constant bias parameters}. Within the realm of applicability of the bias model, we should then find comparable $P_{\rm err}(k)$ among different choices of $k_{\rm max}$. The purpose of this appendix is to investigate how our results change if we diverge from the fiducial scale $k_{\rm max} = 0.4 \ihmpc$ adopted throughout the work. \par 
We take a subset of our HOD samples, the \redmagic DES sample that possesses the highest degree of super-Poisson stochasticity and the DESI sample that showed the most sub-Poisson behavior. We pick $z=0.55$ and $z=0.25$ as benchmarks. This allows us to assess both super-/sub-Poissonianity at low-redshifts where we expect linear bias to be sufficient and intermediate-redshifts where higher-order operators become more important. We perform measurements of the error power spectra using several different $k_{\rm max}$, namely $k_{\rm max} = [0.1, 0.15, 0.2, 0.25, 0.3, 0.35, 0.4, 0.45, 0.5, 0.6, 0.7, 0.75, 1.0]$ which spans quasi-linear scales down to deeply nonlinear scales beyond the expected smallest scale where a perturbative bias expansion is applicable. \par
\begin{figure}
    \centering
    \includegraphics[width=\columnwidth]{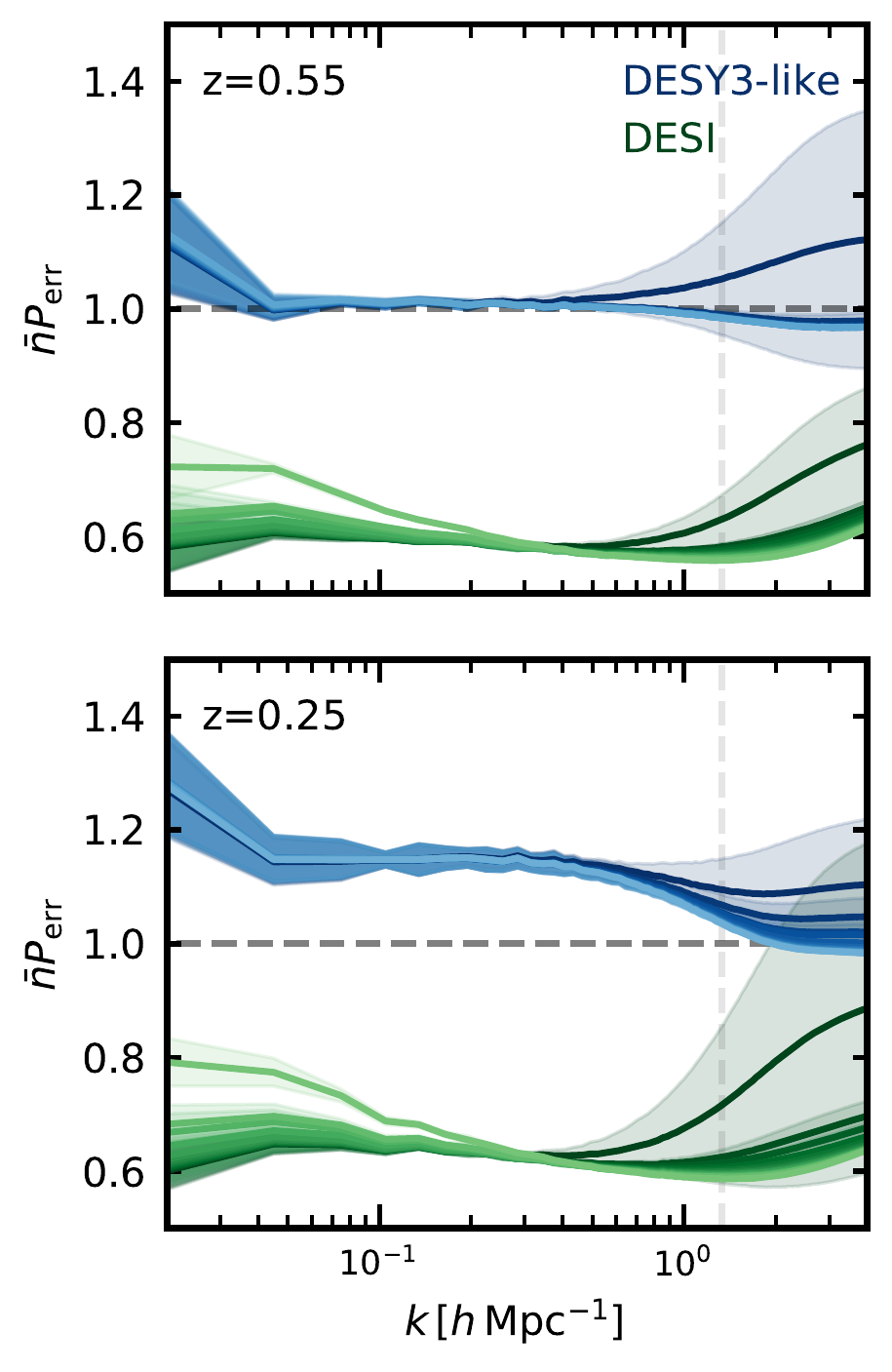}
    \caption{The change in the error power spectrum of the DESI and \redmagic-like samples as we vary the cut-off $k_{\rm max}$ at used to estimate the bias parameters. Cutoffs correspond from $k_{\rm max} = 0.1 \ihmpc$ to $k_{\rm max} = 1.0 \ihmpc$. The lightest (darkest) shade of the color corresponds to the largest (smallest) $k_{\rm max}$ used. }
    \label{fig:kstarvary}
\end{figure}

 
We show our results in the two panels of Fig.~\ref{fig:kstarvary}. For the DES sample we observe that for both redshifts under consideration, the large-scale spectra are very similar independent of $\kmax$. However, when lower $\kmax$ cutoffs are adopted we see a large amount of sample variance in the measured error power spectra at $k \gtrsim 0.4$ which is subsequently reduced when smaller-scale information is incorporated. Notably, including smaller-scale information leads to small-scale power spectra that asymptote to a value close to the Poisson shot-noise prediction. \par 
For the case of the DESI curves, we observe a similarly large scatter at low $\kmax$. However, unlike the DESY3-like sample as we include smaller scale information we eventually reach a regime where the large-scale fit is degraded and the error power spectra are no longer consistent with a constant at large scales, indicating a potential break-down of the bias model. \par 
Nevertheless, we note that the observed $\bar{n}P_{\rm err}$ are quite stable to the choice of $\kmax$ at large scales until extreme values are chosen. This indicates that our fiducial choice of $\kmax = 0.4 \ihmpc$ is adequate for the analysis carried out in this work. 
\section{Covariance between bias parameters}
\label{appendix:C}
\begin{figure*}
    \centering
    \includegraphics[width=\textwidth]{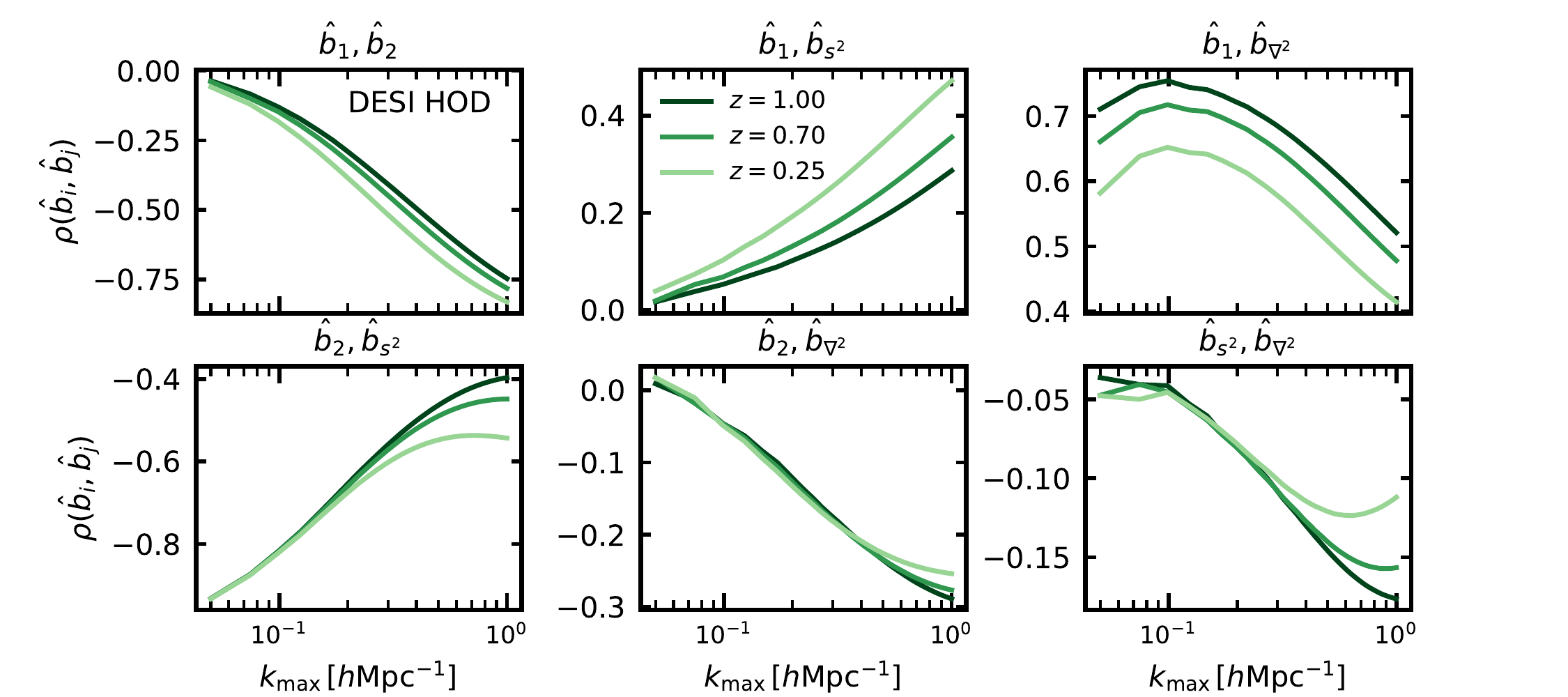}
    \caption{Cross-correlation coefficients for Lagrangian bias parameters from the field-level estimator developed in this work, as a function of $\kmax$, for the case of the DESI HOD across three snapshots. We show the mean curve for the five simulations at our fiducial cosmology, but note that the scatter in the correlation coefficients is negligible. }
    \label{fig:biascorr}
\end{figure*}
In \S~\ref{subsec:Perrstimate} we derived an estimator for the bias parameters at the field level by minimizing the variance of the residual field $\epsilon (\bx) = \delta_h (\bx) - \delta(\bx) - \sum_i b_i \mathcal{O}_i (\bx)$ and commented on the fact the parameters are correlated. In this appendix we derive the co-variance of the bias parameters and report the structure of the correlation matrix as a function of $\kmax$. \par 
The covariance ${\rm Cov}(\hat{b}_i,\hat{b}_j)$ will be given by
\begin{align}
   {\rm Cov}(\hat{b}_i,\hat{b}_j) &= \langle (\hat{b}_i - b_i) (\hat{b}_j - b_j) \rangle  \\
   &= \langle \hat{b}_i \hat{b}_j \rangle - b_i b_j,
\end{align}
where've used the fact that $\langle \hat{b}_i \rangle = b_i$\footnote{This is true under the assumption the stochastic residual field $\epsilon$ and the HEFT component fields $\mathcal{O}_i$ are uncorrelated.}. Recall that from our estimator $\hat{b}_i = M_{ij}^{-1} A_j$ where $M_{ij}$ and $A_j$ are given by Eqns.~\ref{eqn:mij} and \ref{eqn:aj} respectively. First, we note that we may re-write $A_j$ as 
\begin{align}
    A_j &= \int\limits_{\mathclap{|\bk| < k_{\rm max}}} \frac{d^3k}{(2\pi)^3} \mathcal{O}_j (\bk) [\delta_h - \delta_m]^*(\bk), \\
    &= \int\limits_{\mathclap{|\bk| < k_{\rm max}}} \frac{d^3k}{(2\pi)^3} \mathcal{O}_j (\bk) \left [ \epsilon + \sum_i b_i \mathcal{O}_i \right ]^*(\bk),\\
    &\equiv \int_k \mathcal{O}_j  \left [ \epsilon + b_i\mathcal{O}_i \right ]^*,
\end{align}
where in the last line we have introduced a notational convenience to not clutter subsequent equations. The co-variance term is then given by
\begin{align}
&\langle \hat{b}_i \hat{b}_j \rangle = \left \langle M_{ik}^{-1}M_{jn}^{-1}A_k A_n \right \rangle \\
&= \left \langle M_{ik}^{-1}M_{jn}^{-1} \int_{k, k'} \mathcal{O}_k\mathcal{O}_n  \left [ \epsilon + b_i\mathcal{O}_i \right ]^*  \left [ \epsilon + b_i\mathcal{O}_i \right ]^* \right \rangle \\
&= b_i b_j + \left \langle M_{ik}^{-1}M_{jn}^{-1} \int_{k, k'} \mathcal{O}_k\mathcal{O}_n \epsilon \epsilon \right \rangle.
\end{align}
The Hessian $M_{ij}$ is fixed for a given realization, and thus we can remove it from the expectation value. The $b_i b_j$ terms will cancel, and we are left with
\begin{align}
 {\rm Cov}(\hat{b}_i,\hat{b}_j) = M_{ik}^{-1}M_{jn}^{-1} \int_{k,k'} \left \langle \mathcal{O}_k (\bk) \mathcal{O}_n^* (-\bk') \epsilon(\bk) \epsilon^* (-\bk') \right \rangle.
\end{align}
Once again, the component fields $\mathcal{O}_k$ are deterministic given a dark matter realization and so we find that our end result is 
\begin{align}
 {\rm Cov}(\hat{b}_i,\hat{b}_j) = M_{ik}^{-1}M_{jn}^{-1} \int\limits_{\mathclap{|\bk| < k_{\rm max}}} \frac{d^3k}{(2\pi)^3} \mathcal{O}_k (\bk) \mathcal{O}_n^* (\bk) P_{\rm err}(k).
\end{align}
Under the approximation the next-to-leading non-Poisson corrections to $P_{\rm err}(k)$ will be negligible up to $k_{\rm max}$ we then find
\begin{align}
 {\rm Cov}(\hat{b}_i,\hat{b}_j) \approx \frac{M_{ij}^{-1} (\kmax)}{\bar{n}} \left [ a_1 +  O \left (\int_k k^2 \mathcal{O}_k \mathcal{O}_n \right )\right  ].
\end{align}
The correlation coefficient of bias parameters, $\rho_{b_i,b_j}$ is thus given by

\begin{align}
    \label{eqn:biascorr}
    \rho_{\hat{b}_i, \hat{b}_j}(\kmax) = \frac{M_{ij}^{-1}(\kmax)}{\sqrt{M_{ii}^{-1} (\kmax)M_{jj}^{-1}(\kmax)}}.
\end{align}
In Fig.~\ref{fig:biascorr} we show the bias correlation coefficients as computed in Eqn.~\ref{eqn:biascorr} for the DESI HOD across three different redshifts. We observe a pattern of correlations that evolves significantly with $\kmax$ and with redshift. As mentioned in \S~\ref{subsec:Perrstimate}, these correlations can be partially responsible for the running observed in the bias parameters with $k_{\rm max}$. Better quantifying these correlations and their impact in field-level bias parameter estimation is an important step in using the techniques developed in this work to also place priors in the bias parameters themselves, and not just the stochasticity of galaxy samples.

\section{Subsets of operators and $\bar{n}P_{\rm err}$}
\label{appendix:D}
Previous studies of halo stochasticity have resorted to studying the error power spectrum under the assumption of only including linear bias \citep{Hamaus_2010, Baldauf_2013}. The impact of not including higher-order bias operators then manifests itself as a super-Poisson contribution beyond the one-halo enhancement we have previously discussed. \par
\begin{figure*}
    \centering
    \includegraphics[width=0.75\textwidth]{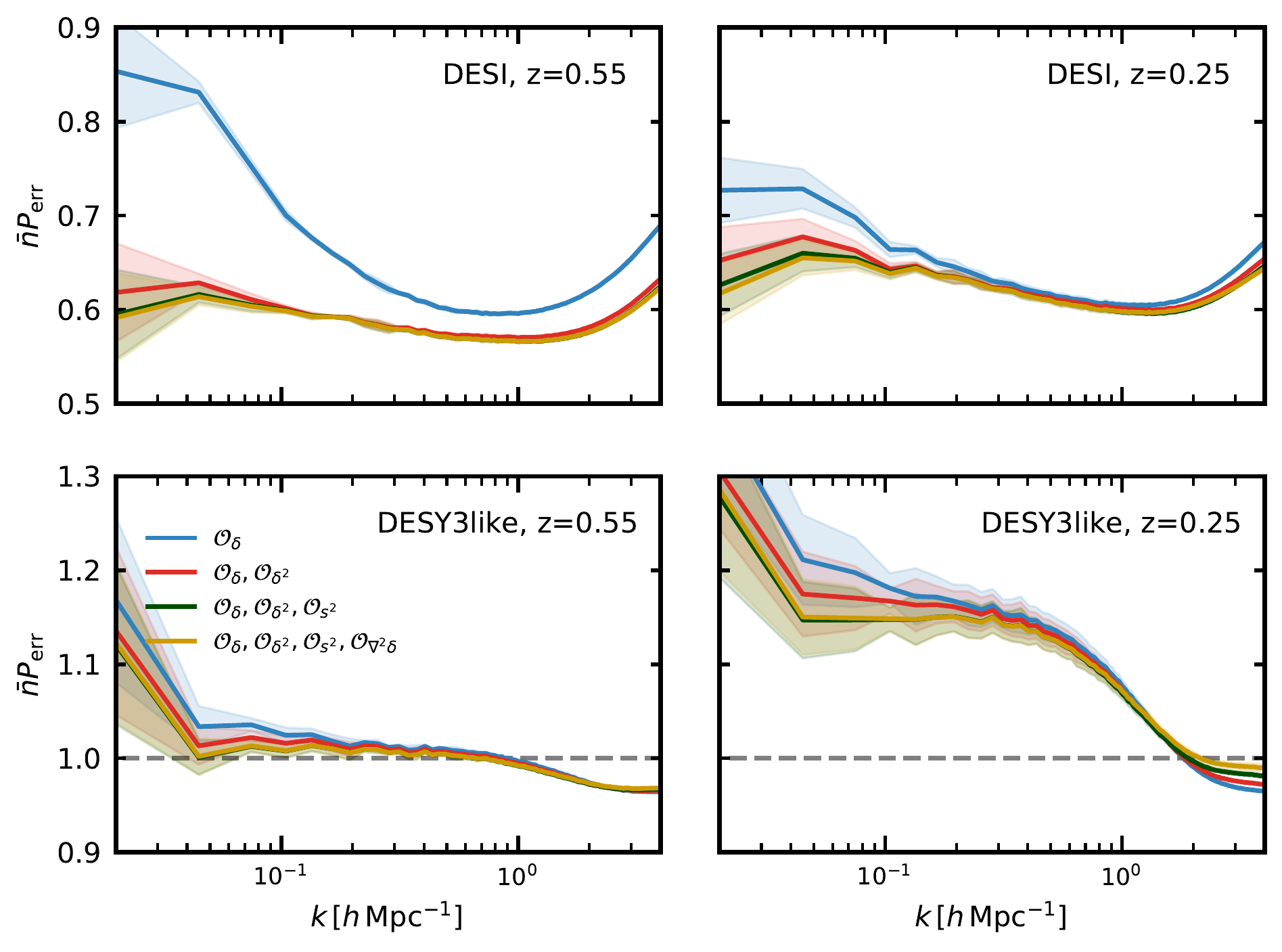}
    \caption{The impact of including higher-order bias operators in the field-level description of different HOD samples, at different snapshots. Each row designates a different type of mock galaxy, while each column indicates the simulation snapshot at which we have carried out these measurements. Note the reduced y-axes compared to other error power spectra presented in this publication. All fits are made using the same fiducial cut-off $k_{\rm max} = 0.4\, h {\rm Mpc}^{-1}$.}
    \label{fig:nbiasplot}
\end{figure*}

 
We investigate the dependence of our $\bar{n}P_{\rm err}$ measurements on including the full set of second-order Lagrangian bias fields. We fit bias parameters to the DESI and \redmagic HOD fields at redshifts $z=0.25$ and $z=0.55$, which allows us to probe the impact of including subsequent operators as a function of redshift. These measurements are reported in Fig.~\ref{fig:nbiasplot}, where we show the impact of including subsequent operators in the error power spectra. \par 
For the DESI sample, we see that the including additional bias operators has the largest effect at large scales. The biggest impact comes from including the quadratic bias operator $\mathcal{O}_{\delta^2}$, which eliminates a significant portion of the excess super-Poissonian stochasticity compared to the expected asymptotic value coming only from exclusion. While the impact of including the tidal shear and non-local bias operators is more modest, their inclusion trends toward flattening the large-scale power spectrum. We also note that the higher redshift snapshot has a significantly larger impact from including additional bias operators. This is consistent with the fact that galaxy samples are more biased tracers of the matter density field at higher redshifts. At very small scales, including additional bias parameters has little effect. \par 
In the case of \redmagic, we find that the impact of including subsequent bias operators is smaller than for the DESI sample. At large scales the trend is to flatten the spectra, but the impact of any given operator is seemingly more modest. We additionally find, for the low-redshift snapshot, that including the higher-derivative bias operator leads to high-$k$ spectra that are closer to the Poisson expectation. A potential explanation is that the excess one-halo enhancement seen in \redmagic can contribute with an amplitude comparable to higher-order operators, making their impact less significant. However, we caution this is a potential explanation and a more cautious investigation is reserved for future work. \par 

\begin{figure*}
    \centering
    \includegraphics[width=\textwidth]{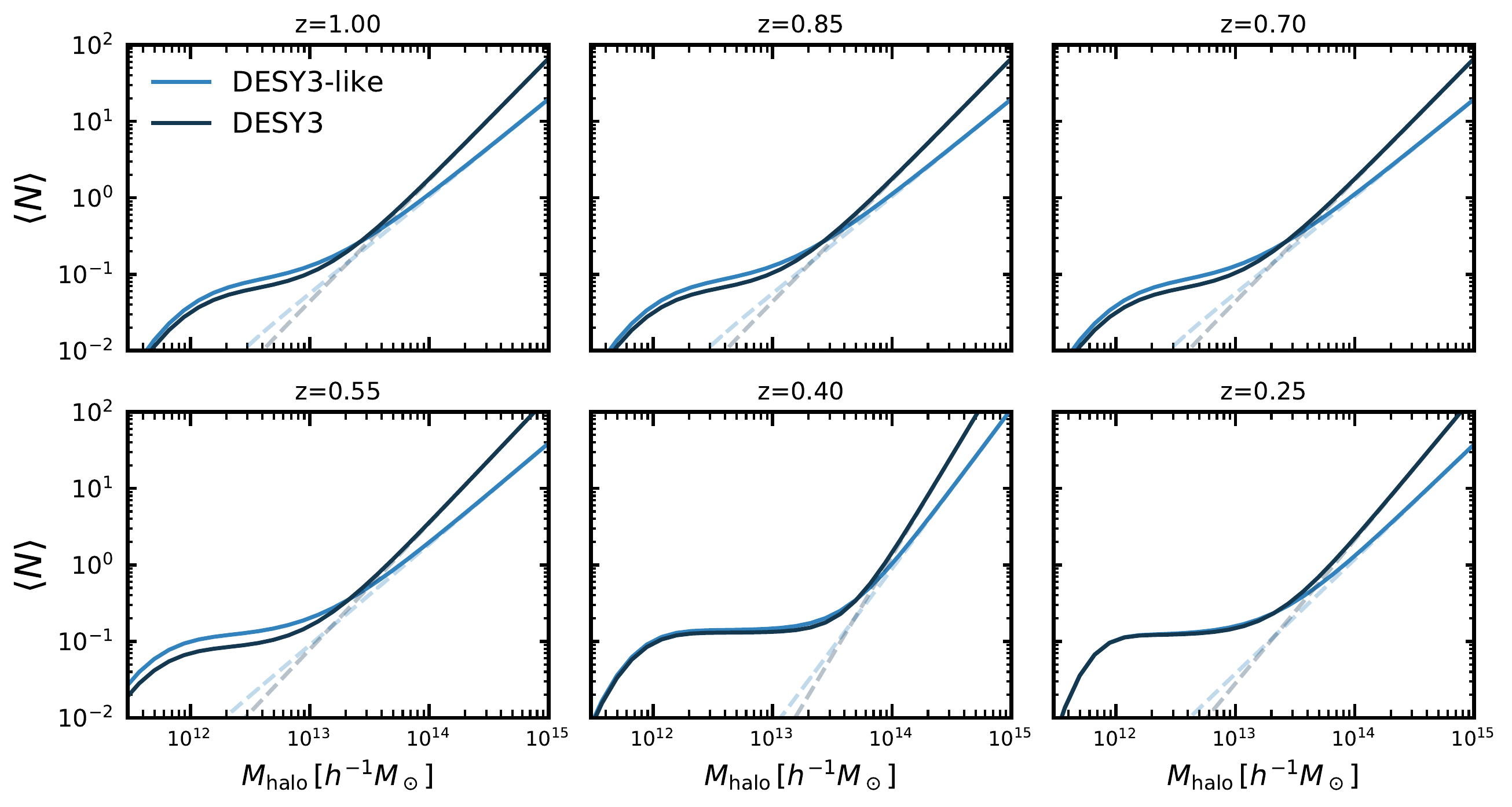}
    \caption{Visualizations of the halo occupation distributions for the DES Y3 \redmagic galaxy sample. The light blue shade corresponds to the `DESY3-like' sample we adopt as our fiducial in this publication. The darker blue shade corresponds to the occupation as constrained in \protect\cite{zacharegkas2021dark}. The solid (dashed) lines show the number of total (satellite) galaxies per halo mass, respectively.}
    \label{fig:desappendix}
\end{figure*}
\end{document}